 \documentclass[aip,jcp,reprint,amsmath,amssymb]{revtex4-1}

 \usepackage{graphicx}
 \usepackage{bm}
 \usepackage{comment}
 \usepackage{epsfig}

 \newcommand{\be}{\begin{equation}} 
 \newcommand{\ee}{\end{equation}}
 \newcommand{\bea}{\begin{eqnarray}} 
 \newcommand{\eea}{\end{eqnarray}}

\begin{document}
 \title{Monte Carlo methods for estimating depletion potentials in highly size-asymmetrical hard sphere mixtures}

 \author{D.~J. Ashton}
 \affiliation{Department of Physics, University of Bath, Bath BA2 7AY, U.K. }
 \author{V.~S\'{a}nchez-Gil}
 \affiliation{Instituto de Qu\'{i}mica F\'{i}sica Rocasolano, CSIC, Serrano 119, E-28006 Madrid, Spain}
 \author{N.~B. Wilding}
 \affiliation{Department of Physics, University of Bath, Bath BA2 7AY, U.K. }

 \begin{abstract}

 We investigate Monte Carlo simulation strategies for determining the
 effective (``depletion'') potential between a pair of hard spheres
 immersed in a dense sea of much smaller hard spheres. Two routes to
 the depletion potential are considered. The first is based on
 estimates of the insertion probability
 of one big sphere in the presence of the other; we describe and
 compare three such methods. The second route exploits collective
 (cluster) updating to sample the depletion potential as a function of
 the separation of the big particles; we describe two such
 methods. For both routes we find that the sampling efficiency at high
 densities of small particles can be enhanced considerably by
 exploiting `geometrical shortcuts' that focus the computational
 effort on a subset of small particles. All the methods we describe
 are readily extendable to particles interacting via arbitrary
 potentials.

 \end{abstract}

 \maketitle

 \section{Introduction}

 Effective potentials arise in theories of complex multicomponent
 fluids such as colloidal suspensions or polymer solutions which
 comprise mixtures of big and small particles. For such a system one
 seeks to integrate out from the full Hamiltonian the degrees of
 freedom of the small particles in order to obtain an effective
 Hamiltonian for the big particles. The motivation for doing so is to
 create an approximate, yet analytically tractable description of the
 true system in terms of a single component model of big particles.
 Unfortunately, obtaining the full effective Hamiltonian is a tall
 order \cite{Likos:2001fk,Belloni2000}. A first step in any
 theoretical treatment is therefore to determine the two-body
 effective potential between a single pair of the big particles in a
 sea of the smaller species.  However, even this task is challenging
 when there exists a very large disparity in size between the
 particles as is common in suspensions containing a mixture of two
 sterically-stabilized colloid species. Such systems are often
 modelled as a highly size-asymmetric binary mixture of hard spheres,
 for which the effective interactions arise from the celebrated
 depletion mechanism \cite{Lekkerkerker:2011}.  As well as being a key
 ingredient in determining effective Hamiltonians for asymmetrical
 hard sphere mixtures, depletion interactions  can be directly measured
 in experiments \cite{Crocker1999}.

 For a pair of big hard spheres in a sea of small hard spheres, the
 effective potential takes the form

 \be 
 \phi_{\rm eff}(r_{\rm bb}) = \phi_{\rm bb}(r_{\rm bb}) + W(r_{\rm bb}), 
 \ee 
 where $\phi_{\rm bb}(r_{\rm bb})$ is the bare hard sphere potential
 between two big spheres of diameter $\sigma_b$ whose centers are
 separated by a distance $r_{\rm bb}$, and $W$ is the ``depletion
 potential'' which is mediated by the small spheres of diameter
 $\sigma_{\rm s}$.  In this paper we consider additive hard sphere
 mixtures so that the big-small interaction diameter $\sigma_{\rm
   bs}=(\sigma_{\rm b}+\sigma_{\rm s})/2$. In that case the depletion
 potential is attractive for small separations of the big spheres, but
 decays in an exponentially damped oscillatory fashion at large
 separations. The physics of the attraction is well understood: The
 exclusion or depletion of the small spheres as the big ones come
 close together results in an increase in free volume available to the
 small species leading to a net increase of entropy
 \cite{Lekkerkerker:2011}.

 A number of theoretical prescriptions exist for determining effective
 potentials, including integral equations (as summarized in the recent
 article by Bo\c{t}an {\em et al}\:\cite{Botan2009}), DFT (see the
 summary in Ashton et al \cite{Ashton:2011kx}) and morphometric theory.
 \cite{Oettel2009,Botan2009} However, these theoretical treatments involve
 approximations, the validity of which need to be checked. Computer
 simulation potentially provides a route to estimating effective
 potentials which is in principle exact, and can therefore be used to
 verify theoretical predictions.  Unfortunately, it too finds the
 regime of large size asymmetry extremely challenging.  The difficulty
 stems from the slow relaxation of the big particles caused by the
 presence of the small ones. Specifically, in order to relax, a big
 particle must diffuse a distance of order its own diameter
 $\sigma_b$. However, for small size ratios,
 $q\equiv\sigma_s/\sigma_b$ and even at quite low volume fractions of
 small particles, very many small particles will typically occupy the space
 surrounding a big particle and these hem it in, greatly hindering its
 movement. In computational terms this issue mandates a very small
 Molecular Dynamics timestep in order to control integration errors,
 while in basic Monte Carlo (MC), a very small trial step-size must be used
 in order to maintain a reasonable acceptance rate. Consequently, the
 computational cost of simulating highly size asymmetric mixtures by
 traditional means is prohibitive at all but very low volume fractions
 of small particles.

 Owing to these difficulties, most previous simulation studies of
 highly size asymmetrical ($q\lesssim0.1$) hard sphere mixtures
 \cite{Biben1996,Dickman1997,Goetzelmann1999,Herring07} have
 adopted an indirect route to measuring depletion potentials based on
 measurements of interparticle {\em force}.  The strategy rests on the
 observation that the force between two big particles can be expressed
 in terms of the contact density of small particles at the surface of
 the big ones \cite{Attard1989,Dickman1997}. By measuring this
 (angularly dependent) contact density for fixed separation $r_{\rm
   bb}$ of the big particles and repeating for separations ranging
 from contact, $r_{\rm bb}=\sigma_{\rm b}$, to $r_{\rm bb}=\infty$,
 one obtains the force profile. This can in turn be
 integrated to yield an estimate of the depletion potential. However, the statistical quality of the data obtained via
 this route is typically quite low, particularly at small $q\leq 0.1$ and
 high densities of small particles. This presumably reflects the
 difficulties of measuring contact densities accurately (which entails
 the extrapolation of data accumulated away from contact) and the
 errors inherent in numerical integration.

 Only a few studies have attempted to measure the depletion potential
 directly for $q\le 0.1$ --see Malherbe and
 Amokrane~\cite{Malherbe2001} for a hard sphere study and Luijten and
 coworkers~\cite{Liu2005,Barr2006} for more general potentials.
 These studies deployed a cluster algorithm (to be described in
 sec.~\ref{sec:gca}) to deal with the problem of slow relaxation
 outlined above. However, this algorithm is limited in the range of
 particle volume fractions for which it will operate efficiently and
 thus there is a need for alternative approaches that extend this range to
 higher values.

An additional drawback of previous studies is that they have treated the
small particles canonically rather than grand canonically. Doing so
complicates comparison with theoretical studies which are typically
formulated in terms of an infinite reservoir of small particles. It is
also at variance with the common experimental situation of a depletant
that is in equilibrium with a bulk reservoir.

 In what follows we consider how Monte Carlo simulation can be used to
 obtain direct and accurate estimates of the depletion potential between two big
 hard spheres separated by a distance $r_{\rm bb}$ immersed in a dense sea
 of small particles at size ratio $q=0.1$. Our focus is on the
 range of available techniques, their implementation and their
 relative utility; comparisons with theoretical predictions have
 appeared elsewhere \cite{Ashton:2011kx}.

 \section{System setup}
 \label{sec:system}

\begin{figure}[h]
   \includegraphics[type=pdf,ext=.pdf,read=.pdf,width=0.95\columnwidth,clip=true]{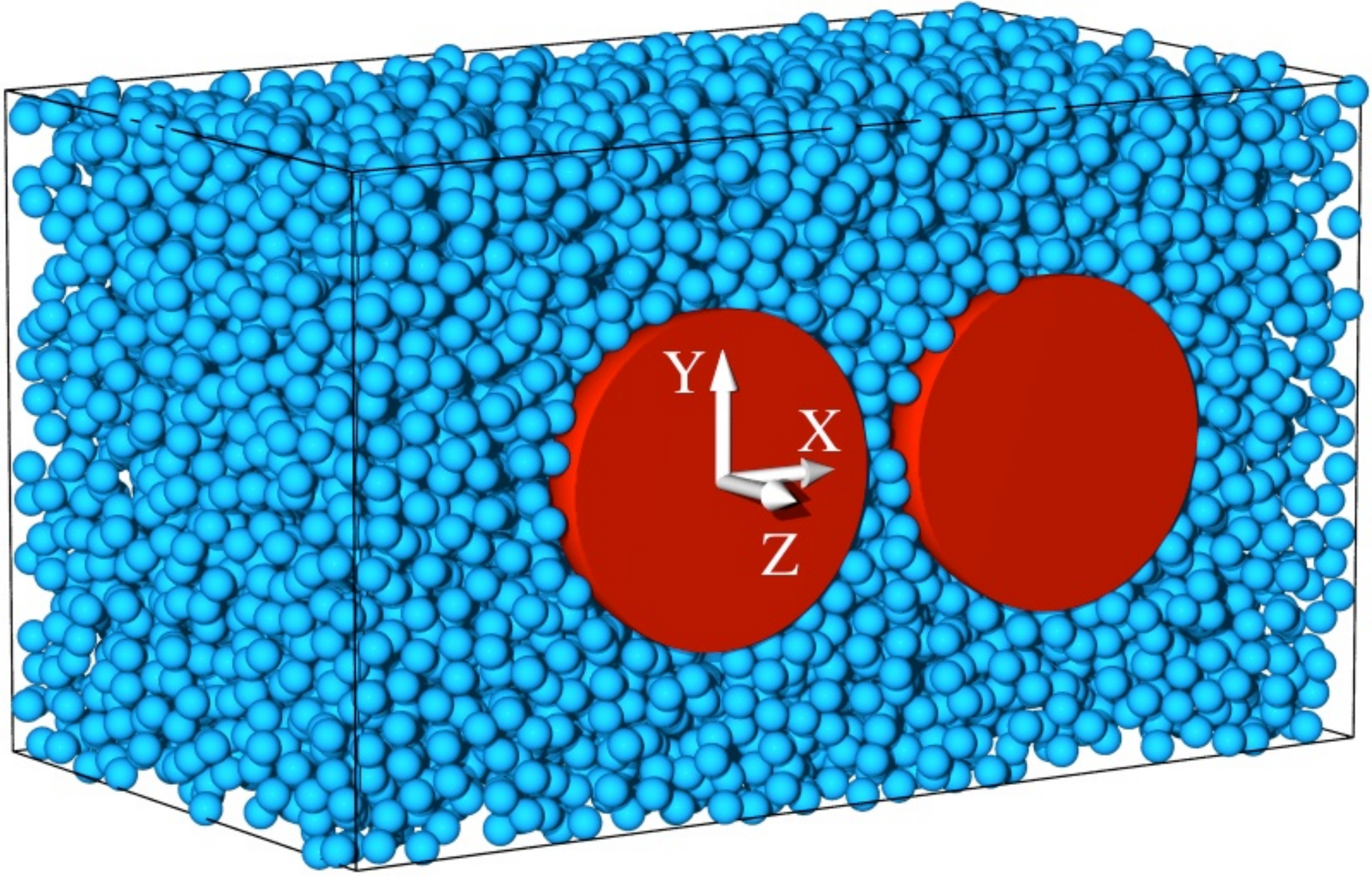}

 \caption{A cross section through a snapshot of a configuration as
   described in the text. The simulation box contains a pair of big
   hard spheres, one of which is fixed at the origin, while the other
   is located at $x=r_{\rm bb}, y=0, z=0$, with $r_{\rm bb}=1.19$ in
   this case. The big particles are in equilibrium with a fluid of
   small hard spheres (size ratio $q=0.1$) at reservoir volume
   fraction $\eta_{\rm s}^r=0.32$. The section shown corresponds to the
   region $z<0$.}

   \label{fig:setup}
 \end{figure}

 The simulation setup that we consider for the measurement of
 depletion potentials is depicted in cross section in
 Fig.~\ref{fig:setup}. It comprises a cuboidal periodic simulation box
 with dimensions $L_x=3.5\sigma_{\rm b}, L_y=L_z=2.0\sigma_{\rm
   b}$. This box accommodates two big hard spheres and a large number of
 small ones. (Though as described below, in some instances it will prove beneficial to take one
 of the big particles to be a hard shell).  Owing to the spherical symmetry
 of the depletion potential we can, without loss of generality, fix
 the center of one of the big particles at the origin, while
 constraining the center of the other to occupy points along the
 $x$-axis at $x=r_{\rm bb}\ge\sigma_{\rm b}$. The only exception to
 this arrangement is the cluster algorithm to be discussed separately
 in Sec.~\ref{sec:gca}.

We set the size of the small particles to be $\sigma_{\rm
  s}=0.1\sigma_{\rm b}$, ie. $q=0.1$. We also elect to treat them
grand canonically so that their total number fluctuates. Conceptually
this corresponds to a colloidal system connected to a reservoir of
depletant particles whose properties are parameterized in terms of
either the reservoir volume fraction $\eta_{\rm s}^r=\pi\rho_{\rm
  s}\sigma_{\rm s}^3/6$, (with $\rho_{s}=N_{\rm s}/V$ the reservoir
number density) or equivalently the conjugate chemical potential
$\mu_s^r$. In practical terms, use of the grand canonical ensemble
aids relaxation of small particle configurations because particle
transfers (insertions and deletions) can be performed very
efficiently. However to utilize this ensemble one needs to know
accurately the chemical potential corresponding to a given
$\eta_{\rm s}^r$. We obtain this from the equation of state of Kolafa {\em et al}\;
 \cite{Kolafa:2004vn}, which we have checked provides a highly
accurate representation of grand canonical ensemble simulation
data. Transfers of small particles are effected using a standard grand
canonical approach \cite{Frenkelsmit2002}. For the most part we
consider the case of a rather high reservoir volume fraction of small
particles, $\eta_{\rm s}^r=0.32$, which also corresponds to the conditions
depicted in the configurational snapshot of Fig.~\ref{fig:setup}.

 \section{Overview of computational strategies}
 \label{sec:overview}

We shall investigate two distinct routes to obtaining estimates of
depletion potentials which we outline here before going into
detail in Secs.~\ref{sec:gradual} and ~\ref{sec:cluster}. The first
route is based on measurements of the insertion probability of one big
sphere in the presence of the other; the second is based on direct
sampling of free energy differences associated with variations in the
separation between the two big spheres.

 \subsection{Insertion route and the shell trick}
 \label{sec:chempotroute}

Let $\mu_{\rm ex}(r_{\rm bb})$ be the excess chemical potential
associated with inserting a big sphere at some prescribed distance
$r_{\rm bb}$ from another big sphere. It is straightforward to show
that this function is equivalent to the effective potential up to an
additive constant \cite{Mladek2011,Ashton:2011kx} i.e.

 \be
 W(r_{\rm bb})=\mu_{\rm ex}(r_{\rm bb})-C,
 \label{eq:effpot}
 \ee
 where the constant 

 \be
  C=\lim_{r_{\rm bb} \to{\infty}}{\mu_{\rm ex}(r_{\rm bb})}.
 \label{eq:const}
 \ee

 To facilitate estimates of the excess chemical potential, one can appeal to the Widom insertion formula
 \cite{Widom:1963zr}, which in the case of hard particles reads

 \be
 \mu_{\rm ex}(r_{\rm bb})=-\beta^{-1}\ln \left [p_{i}(r_{\rm bb})\right]\:.
 \label{eq:mu_ins}
 \ee
 Here $p_i(r_{\rm bb})$ is the probability that an attempt to insert a
 big particle at $x=r_{\rm bb}$ incurs no overlaps with small
 particles; it is calculated with respect to the ensemble of
 configurations of the small particles. $\beta$ is the inverse
 temperature, which in hard particle systems simply serves to bestow
 free energies with the appropriate dimensions; accordingly we shall henceforth set it to unity.

It follows from Eqs.~\ref{eq:effpot}-\ref{eq:mu_ins} that the depletion potential can be expressed in terms of
insertion probabilities as

 \be
  W(r_{\rm bb})=\ln\left(\frac{p_{i}(\infty)}{p_{i}(r_{\rm bb})}\right)\:,
 \label{eq:effpot_ins}
 \ee
 where $p_{i}(\infty)$ represents the insertion probability for infinite
 separation of the big spheres, which in practical terms can be
 determined as the insertion probability of a big sphere in
 a simulation box containing only small particles.

\begin{figure}[h]

   \includegraphics[type=pdf,ext=.pdf,read=.pdf,width=0.95\columnwidth,clip=true]{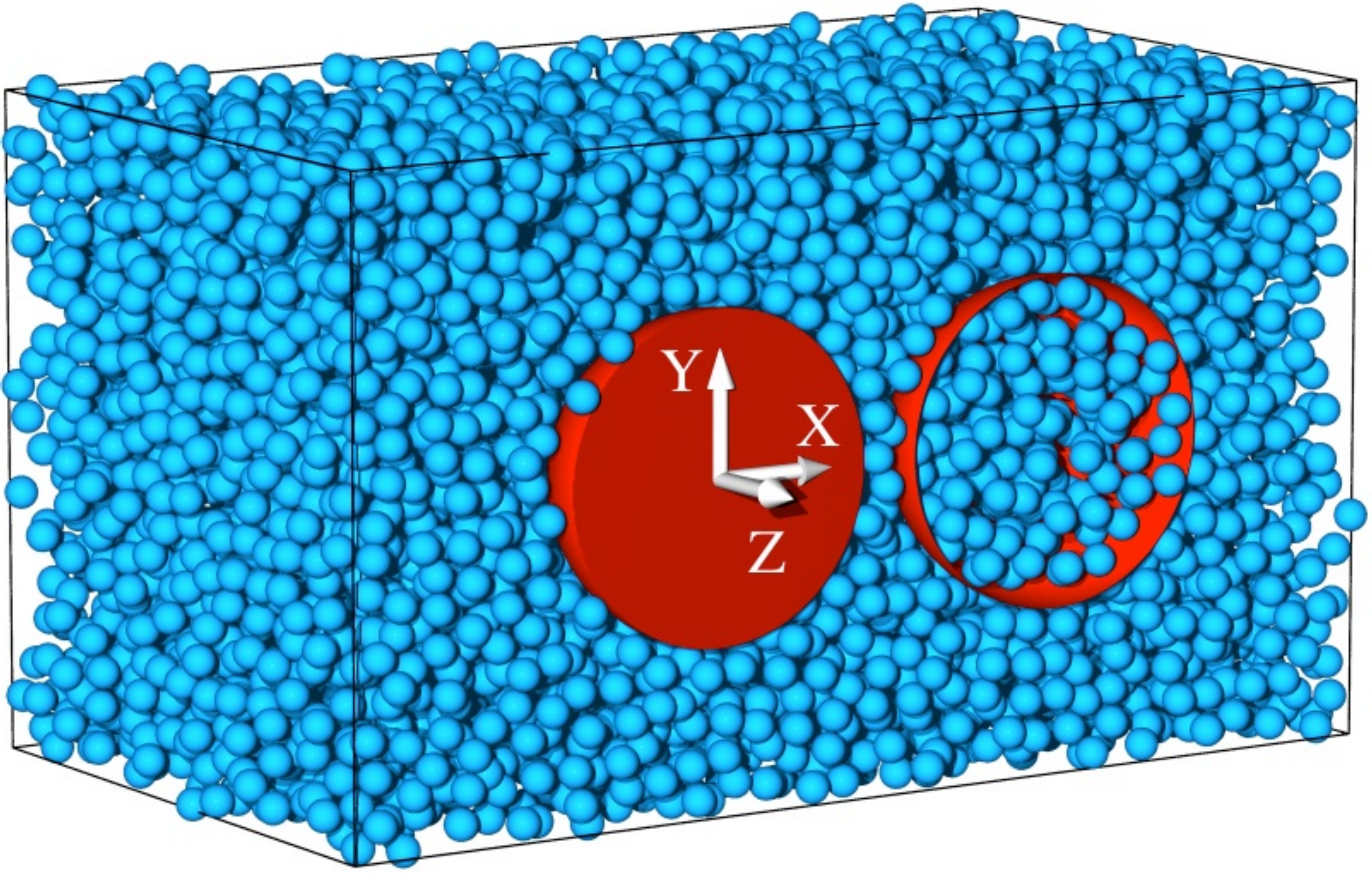}

    \caption{A cross section through a configuration containing a big
      hard sphere fixed at the origin (left) and a fully inserted big hard shell (right)
      in equilibrium with a fluid of small particles at
      $\eta_{\rm s}^r=0.32$. The particle size ratio is $q=0.1$ and the
      separation is $r_{\rm bb}=1.19$. The section shown corresponds
      to $z<0$.}

   \label{fig:shell}

 \end{figure}

 The computational task is then to measure the insertion probability
 $p_i(r_{\rm bb})$. Unfortunately, for the values of $\eta_{\rm s}^r$
 of interest this probability is almost vanishingly small, a fact
 which renders simple sampling ineffective. Consequently we adopt a
 bespoke `gradual insertion' approach, based on the use of tunable
 interactions and biased Monte Carlo sampling. Details of this approach
 are postponed until Sec.~\ref{sec:gradual}.  Here it suffices to note that in
 implementing such schemes a very useful ``geometrical shortcut''
 derives from the fact that it is not actually necessary to consider
 the insertion probability of a big hard {\em sphere} in order to
 calculate the depletion potential. Instead it is sufficient and
 (generally much more efficient) to measure the insertion probability
 for a hard {\em shell} of diameter $\sigma_b$ having infinitesimal
 thickness, as shown in the snapshot of Fig.~\ref{fig:shell}.  The
 essential observation is that when fully inserted, a hard shell
 particle encloses a number of small particles and although these
 remain in equilibrium with the reservoir (by means of particle
 transfers) they are fully screened from the rest of the system
 because their surfaces cannot penetrate the shell wall. Thus
 the contribution to the partition function from the enclosed
 particles is independent of $r_{\rm bb}$, and therefore represents a
 constant contribution to $\mu^{ex}(r_{\rm bb})$ which vanishes from
 the difference in Eq.~\ref{eq:effpot}. Accordingly
 Eq.~\ref{eq:effpot_ins} applies equally to shell insertion as it does
 to sphere insertion. Of course from a computational standpoint, the
 task of inserting a hard shell is much less challenging than that of
 inserting a hard sphere (as can be appreciated by comparing
 Figs.~\ref{fig:setup} and~\ref{fig:shell}): essentially the insertion
 probability falls with the particle size ratio like $q^2$ rather than
 $q^3$. Shell insertion is deployed in each of the three gradual
 insertion methods to be described in Sec.~\ref{sec:gradual}.

 A further geometrical shortcut results from noting that the
 convergence of the ensemble average over small particle
 configurations required to calculate the shell insertion probability
 depends on how quickly the small particles in the region of the shell
 decorrelate. To enhance this relaxation rate we preferentially
 perform grand canonical insertions and deletions of small particles
 within a shell subvolume of radius $0.7\sigma_b\leq r\leq 1.3\sigma_b$ centered on the
 shell. Updates inside the subvolume occur with a frequency $50$-fold
 that of outside.  This approach --which satisfies detailed
 balance-- greatly reduces the time spent updating small particles
 whose coordinates are relatively unimportant for the quantity we wish
 to estimate.

 \subsection{Direct sampling route}

The particle insertion approach outlined above relies on extracting
the depletion potential from differences in the measured values of the
insertion probability as a function of $r_{\rm bb}$. However, even
when using the shell insertion trick, the difference $\ln
p_{i}(\infty)-\ln p_{i}(r_{\rm bb})$ that provides the depletion
potential via Eq.~\ref{eq:effpot_ins}, is (notwithstanding the
logarithm) typically small compared to the absolute values of $\ln
p_{i}(\infty)$ and $\ln p_{i}(r_{\rm bb})$. Potentially, therefore, a
great deal of computational effort is required to obtain a reasonable
accuracy in $W(r_{\rm bb})$. In view of this we have investigated an
alternative strategy for obtaining the depletion potential which
directly measures {\em changes} in the free energy as the separation
between the two big spheres is varied.  To achieve this, however,
specialist methods are required to overcome the steric hindrance to
the displacement of a big particle in a sea of much smaller ones. In
section~\ref{sec:cluster} we consider two methods that enable such
displacements via collective updates of a big sphere and many small
ones. They are: (i) the cluster algorithm of Dress and
Krauth\cite{Dress1995}, which allows the depletion potential to be
built up directly from the sampled histogram of big particle
separations, and (ii) a new constrained biased cluster move, which
permits estimates of the free energy difference associated with a
prescribed displacement of a big particle.

 \section{Insertion route: Implementations}

 \label{sec:gradual}

 In this section we outline three methods that exploit the insertion
 route to determine the depletion potential. The basic idea is to to
 fix a hard sphere at the origin and then estimate the probability of
 inserting a hard shell at coordinates $x=r_{\rm bb}, y=z=0$. In
 practice, however, for highly size asymmetrical mixtures and at all
 but the smallest values of $\eta_{\rm s}^r$, simple sampling of the
 insertion probability is too inefficient to yield accurate
 results. Instead a more elaborate {\em gradual} insertion technique is
 required to render the approach feasible. We note that key elements of
 the relevant strategies and general sampling issues for determining
 insertion probabilities (and thence excess chemical potentials) have
 been discussed previously elsewhere \cite{Nezbeda1991, Attard1993, wilding1994a,
   Kofke:1997,Bruce2003}, though not in the context of highly size
 asymmetrical fluid mixtures.

 \subsection{Method I: Expanded ensemble}

 \label{sec:methodI}

 This method, which has been briefly reported previously
 \cite{Ashton:2011kx} draws on earlier related studies.
 \cite{Nezbeda1991, Attard1993, wilding1994a, lyubartsev1992} It
 involves defining an extended set of states for the
 interaction between the shell particle and the small
 particles and implementing Monte Carlo updates that make transitions
 between these states.

 \subsubsection{Description}

To estimate $p_i(r_{\rm bb})$ for the shell we suppose that it
can exist in one of $M$ possible `ghost' states or `stages' in
which it interacts with a small hard sphere (a distance $r_{\rm bs}$
away) via the potential

 \begin{equation}
   \phi^{({\rm m})}_{\rm g}(r_{\rm bs})= \begin{cases} -\ln\lambda^{({\mathrm m})},  & \text{$(\sigma_{\rm b}-\sigma_{\rm s})/2<r_{\rm bs}<\sigma_{\rm bs}$}\\
     0, &\text{otherwise.}\end{cases}
 \label{eq:ghost-shell}
 \end{equation}
 Here ${\rm m}=0\ldots M-1$ (an integer) indexes the stages, while the associated
 coupling parameter $0\le \lambda^{({\rm m})}\le 1$ controls the strength of
 the repulsion between the big particle and the small ones. Note that for $\lambda^{({\rm m})}>0$ the
 repulsion is {\em finite} so that overlaps between small particles and the big
 one can occur. If we denote by $N_o$ the instantaneous number of such overlaps, 
 then the configurational energy associated with the shell in stage ${\rm m}$ is

 \be
 \Phi_{\rm g}^{({\rm m})}=-N_o\ln\lambda^{({\rm m})}\:.
 \label{eq:energy}
 \ee
 
Clearly for $\lambda^{({\rm m})}=1$, the shell is completely
non-interacting, while for $\lambda^{({\rm m})}=0$ it is infinitely
repulsive.  To span this range we set the extremal stages
$\lambda^{(0)}=1$ and $\lambda^{(M-1)}=0$ (in fact we choose
$\lambda^{(M-1)}=10^{-9}$ to avoid numerical infinities), and define
a set of $M-2$ intermediate stages $\lambda^{({\rm m})}, {\rm m}=1,\ldots, M-2$ that
facilitate efficient MC sampling over the entire range ${\rm m}=0,\ldots
, M-1$, i.e. that permits the shell interaction to fluctuate smoothly
between the two extremes of interaction strength.

 Details of a suitable Metropolis scheme for sampling the full range
 of $m=0\ldots M-1$ have been described previously.
 \cite{wilding1994a,Ashton:2011fk} The basic idea is to perform grand
 canonical simulation of the small particles, supplemented by MC
 updates that allow transitions ${\rm m}\to {\rm m}^\prime={\rm m}\pm 1$ in the stage. 
 These transitions are accepted or rejected probabilistically on the
 basis of the change in the configurational energy,
 Eq.~\ref{eq:energy}. Specifically

 \begin{equation}
 p_a(m\rightarrow {\rm m}^\prime )={\rm min}\left(1,\exp{[-(\Phi_{\rm g}^{({\rm m}^\prime)}-\Phi_{\rm g}^{({\rm m})})  +  \Delta w]} \right)\:,
 \label{eq:accprob1}
 \end{equation}
 where $\Delta w=w^{({\rm m}^\prime)}-w^{({\rm m})}$, with $w^{({\rm m})}$ a
 prescribed weight associated with stage ${\rm m}$ (see below). Note that for transitions that
 depart from the extremal stages ${\rm m}=0$ or ${\rm m}=M-1$, it is necessary to reject proposals that would take ${\rm m}$ outside the range $(0,M-1)$.

The weights are chosen, as described below, such as to allow the
system to smoothly sample the entire range of ${\rm m}$. Over the course of
a sufficiently long run, the sampling results in the system visiting
all the $M$ stages repeatedly, permitting a histogram $\tilde {H}({\rm m})$ of
their relative probabilities to be accumulated. From this biased
histogram, one unfolds the weight factors to obtain an estimate of the
unbiased histogram:

\be
H({\rm m})=\tilde{H}({\rm m})\exp{(w^{({\rm m})})}\:.
\label{eq:whist}
\ee
After normalizing to unit integrated weight, this histogram provides
an estimate of the relative probability $p({\rm m}|r_{\rm bb})$ of finding
the system in each of the $M$ stages. The insertion
probability is simply the relative probability of finding the system
in the extremal stages:

\be
p_i(r_{\rm bb})=\frac{p(M-1|r_{\rm bb})}{p(0|r_{\rm bb})}\:,
\label{eq:pm}
\ee
from which the effective potential (up to a
constant) follows via Eq.~\ref{eq:effpot_ins}. Repeating the
measurement for a succession of values of $r_{\rm bb}$ allows
construction of the entire depletion potential.

 \subsubsection{Remarks and results}
\label{sec:methodIremarks}

The implementation of method I entails a certain degree of
preliminary work. Firstly one must decide on the number of stages $M$
and their locations in $\lambda\in [0,1)$, ie. the set $\{\lambda^{({\rm m})}\},
  {\rm m}=1\ldots M-2$ of intermediate stages that interpolate between the
  extremal values of  $\lambda^{(0)}=1$ and $\lambda^{(M-1)}=10^{-9}$.  It is
  important that these choices result in MC transitions ${\rm m}\to {\rm m}\pm 1$
  that are approximately equally likely in both directions and have a
  reasonably high rate of acceptance. To achieve this we perform a
  preliminary run in which we consider a single big ghost shell in the
  reservoir of small particles. We initially employ a large set of $1000$
  ghost stages, evenly spaced in $\ln \lambda$, and (in short runs)
  measure the distribution of overlaps $p(N_{\rm o}|\lambda^{({\rm m})})$
  for each. From this set we select a subset of $M$ stages for which
  the acceptance rate for transitions ${\rm m}\to {\rm m}\pm 1$ is approximately
  $20\%$. A convenient basis for this selection is provided by
  Eq.~\ref{eq:zwanzig1} which will be discussed in
  Sec.~\ref{sec:methodII}. Choosing a low acceptance rate leads to a
  smaller required number of stages $M$, while a large
  acceptance rate necessitates a correspondingly larger $M$. Although
  we find empirically that the overall efficiency of the method is not
  particularly sensitive to the choice of acceptance rate (provided it
  lies in the range $10\%-50\%$), the $20\%$ figure that we quote
  seems to strike a reasonable balance between the length of the
  sampling path required to span the $M$ stages and the transition
  rate.

Secondly one needs to prescribe a suitable set of $M$ weights
$\{w^{({\rm m})}\}$ for use in the acceptance probability
Eq.~\ref{eq:accprob1}.  The role of these weights is to bias the
acceptance rates such as to enhance the sampling of states of
low probability. Generally speaking a suitable set of weights is one which
ensures approximately uniform sampling of the $M$ stages
\cite{lyubartsev1992}. The weights can be determined using a variety
of methods, though we favor the Transition Matrix Monte Carlo (TMMC)
method detailed in Appendix~\ref{sec:tmmc}. Note that having
determined a suitable set of weights for one value of the big particle
separation $r_{\rm bb}$, this set will (typically) perform
adequately at all values of $r_{\rm bb}$ to be studied, at least
provided the variations in the depletion potential are not too large,
as is certainly the case for the range $\eta_{\rm s}^r\leq 0.32$
considered here. Similarly one does not have to choose a new set of
$\{\lambda^{(m)}\}$ for each choice of the big particle separation
$r_{\rm bb}$, a single choice performs adequately for all separations.

\begin{figure}[h]
   \includegraphics[type=pdf,ext=.pdf,read=.pdf,width=0.95\columnwidth,clip=true]{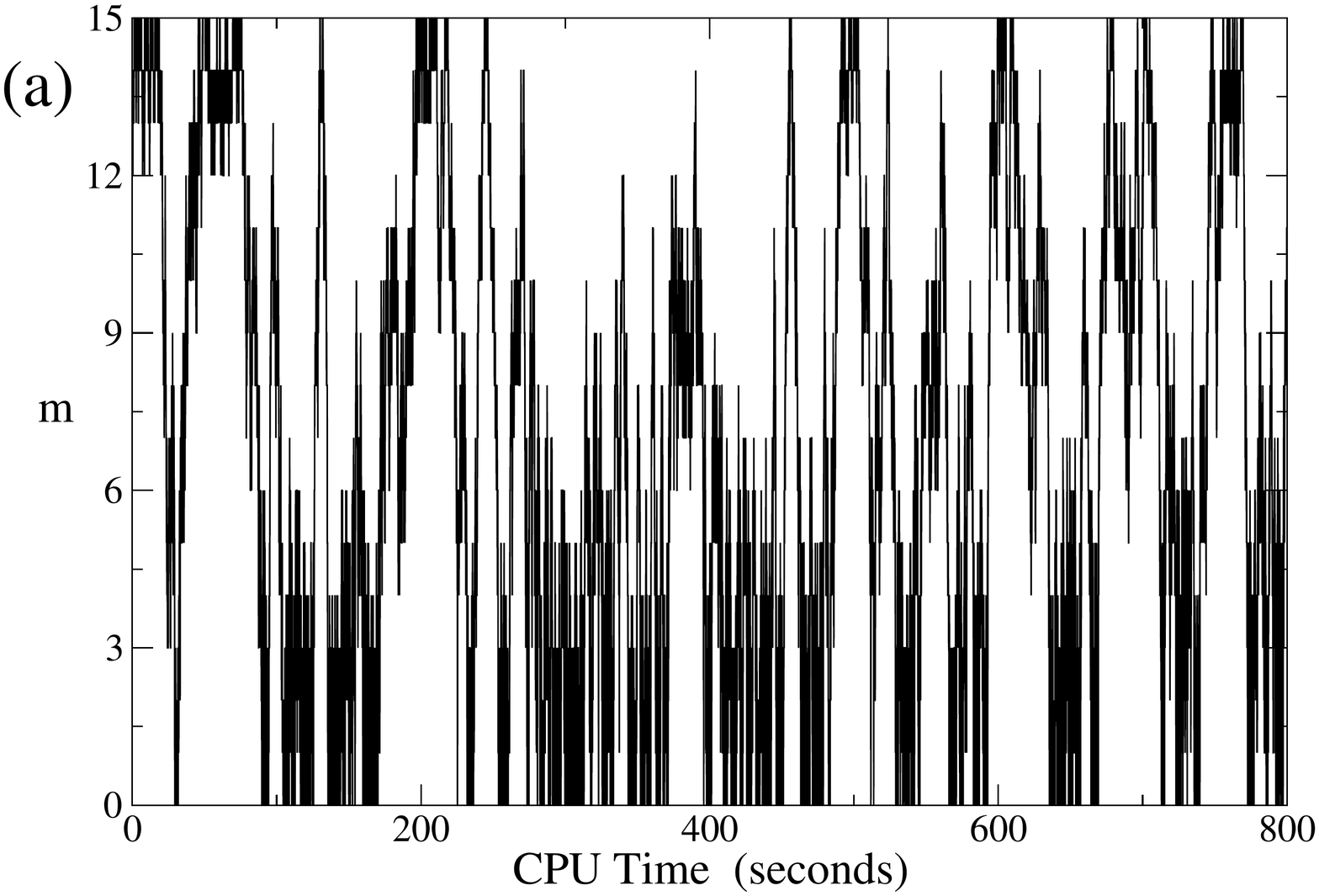}
   \includegraphics[type=pdf,ext=.pdf,read=.pdf,width=0.95\columnwidth,clip=true]{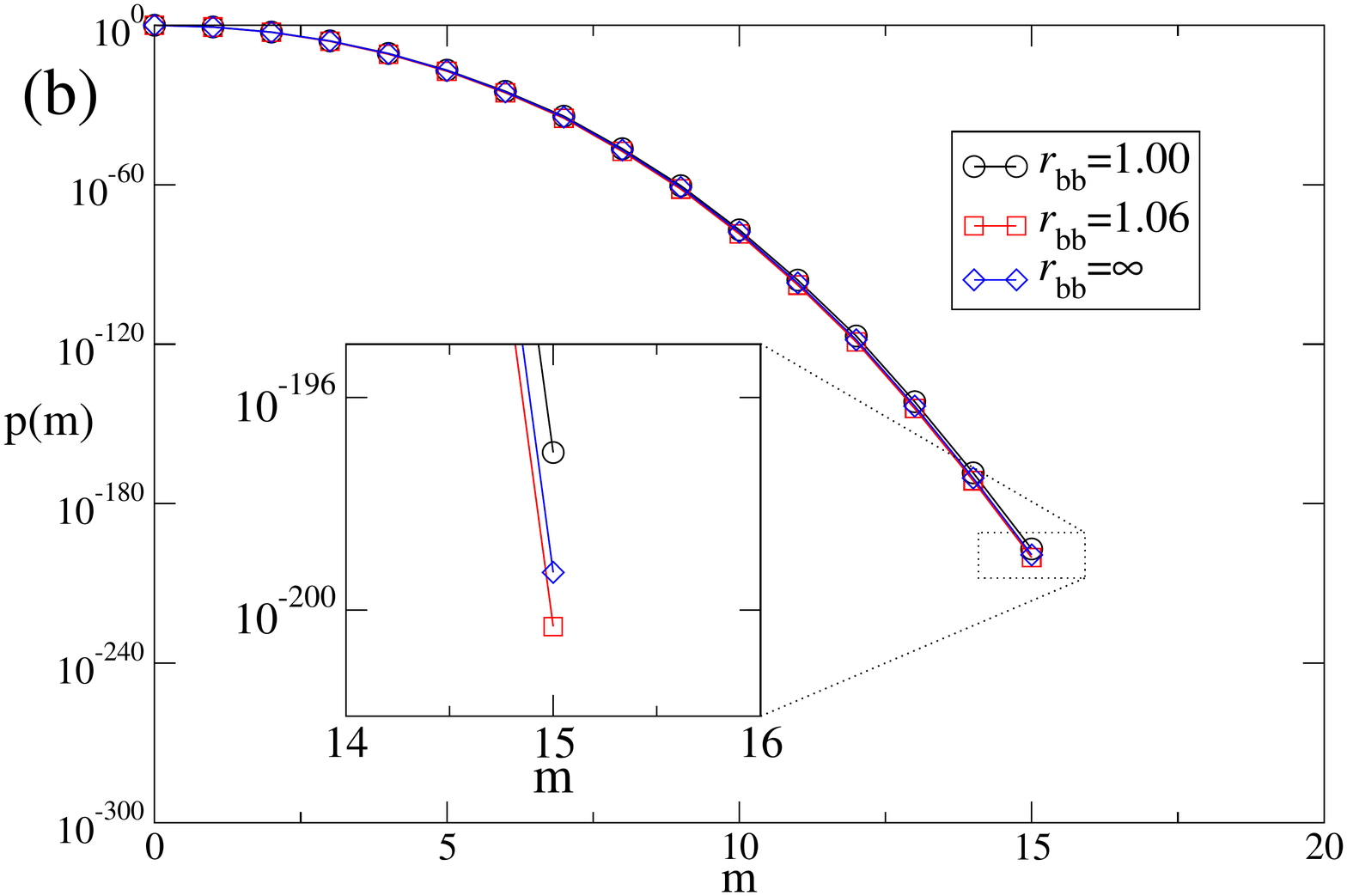}

    \caption{{\bf (a)} $\lambda^{({\rm m})}$ {\em vs} CPU time on a $2$ GHz
      processor for $r_{\rm bb}=1.06$, $\eta_{\rm s}^r=0.32, q=0.1$ as
      obtained for method I. The figure gives an impression of the
      typical time scale required to sample all $M=16$ stages, but constitutes
      only a small portion of the full run which comprised $35$ CPU
      hours. {\bf (b)} the unfolded histogram $p(\lambda^{({\rm
          m})}|r_{\rm bb})$ at a selection of values of $r_{\rm
        bb}$. Differences in the insertion probability
      $p(\lambda^{(M-1)}|r_{\rm bb})/p(\lambda^{(0)}|r_{\rm bb})$
      (inset) provide estimates for the variations in the depletion
      potential.}

   \label{fig:plam}
 \end{figure}

Fig.~\ref{fig:plam} shows data accumulated for $r_{\rm bb}=1.06, q=0.1,
\eta_{\rm s}^r=0.32$. For this state point, $M=16$ stages were
required to realize a $20\%$ acceptance rate for transitions in ${\rm m}$. A
portion of the time series resulting from the sampling of ${\rm m}$ is shown
in Fig.~\ref{fig:plam}(a), giving an impression of the timescale over which the
sampling covers the entire range. The estimates of the probability
distribution $p(\lambda^{({\rm m})}|r_{\rm bb})$ that results from unfolding
the weights from the measured histogram $\tilde H({\rm m})$
(cf. Eqs.~\ref{eq:whist} and ~\ref{eq:pm}) is shown in
Fig.~\ref{fig:plam}(b).  From this, the insertion
probability can be read off directly; it is found to be $O(10^{-200})$,
demonstrating the scale of the depths in probability that the method
allows one to plumb. The rationale for the extreme improbability of successfully
inserting a shell without the support of biased sampling is
to be found in Fig.~\ref{fig:shell}, specifically in the tightness of
the small particle packing at this value of $\eta_{\rm s}^r$.

Finally in this subsection we remark that since the full depletion
potential is built up from separate and independent measurements of
the insertion probability at various values of $r_{\rm bb}$, there is
the opportunity to exploit parallelism by farming out each measurement
on multi-core processors.

 \subsection{Method II: Multiple overlapping histograms}
 \label{sec:methodII}

 Our second approach is related to the previous one in that a set of
 $M$ stages are used to control the strength of
 interaction between the shell and the small particles in the manner
 described by Eq.~\ref{eq:ghost-shell}. The difference is that here we
 don't actually implement transitions $\lambda^{({\rm m})}\to\lambda^{({\rm m}\pm 1)}$,
 instead we simply measure the free energy difference between
 successive values of $\lambda$ via an exact free energy perturbation method.

 \subsubsection{Description}

The relevant expression for calculating free energy differences is the
well known formula of Zwanzig \cite{Zwanzig:1954fk}, which in our
case, for a transition ${\rm m}\to {\rm m}^\prime={\rm m}+1$ reads:

 \bea
 F^{({\rm m}^\prime)}\!- \!F^{({\rm m})} &=& -\ln\left\langle \exp \left[-(\Phi_g^{({\rm m}^\prime)}- \Phi_g^{({\rm m})})\right] \right\rangle_{{\rm m},r_{\rm bb}}\nonumber\\ 
\: &=& -\ln\left\langle \exp \left[N_{\rm o} \ln \frac{\lambda^{({\rm m})}}{\lambda^{({\rm m}^\prime)}} \right] \right\rangle_{{\rm m},r_{\rm bb}}\nonumber\\
 \:      &=& -\ln\left( \sum_{N_{\rm o}} \left( \frac{\lambda^{({\rm m})}}{\lambda^{({\rm m}^\prime)}}\right)^{N_{\rm o}} \!P(N_{\rm o}|\lambda^{({\rm m})},r_{\rm bb})\right).\nonumber\\
 \label{eq:zwanzig1}
 \eea
Here the ensemble average is with respect to the small particle configurations in stage ${\rm m}$, given a big particle separation $r_{\rm bb}$.

We can apply this formula in the forward and reverse directions, averaging the result to find:

 \be
 F^{({\rm m}^\prime)}\!-\! F^{({\rm m})}=\frac{1}{2}\ln \frac{ \sum_{N_{\rm o}}\left(\frac{\lambda^{({\rm m}^\prime)}}{\lambda^{({\rm m})}}\right)^{N_{\rm o}}P(N_{\rm o}|\lambda^{({\rm m}^\prime)},r_{\rm bb})}{\sum_{N_{\rm o}}\left(\frac{\lambda^{({\rm m})}}{\lambda^{({\rm m}^\prime)}}\right)^{N_{\rm o}}P(N_{\rm o}|\lambda^{({\rm m})},r_{\rm bb})}\:.
 \label{eq:zwanzig2}
 \ee
Thus, operationally, having chosen a suitable set of intermediates
$\{\lambda^{({\rm m})}\}$, one simply measures the distribution of overlaps
$P(N_{\rm o}|\lambda^{({\rm m})},r_{\rm bb})$ at each $\lambda^{({\rm m})}$. This
yields the insertion probability via

 \begin{equation}
 \ln p_i(r_{\rm bb})=F^{(0)}-F^{(M-1)}\:.
 \label{eq:whtdiff}
 \end{equation}
 The depletion potential then follows by repeating this measurement for a sequence of values of
 $r_{\rm bb}$ and utilizing Eq.~\ref{eq:effpot_ins} as was done in Sec.~\ref{sec:methodI}.

 \subsubsection{Remarks and results}

 For this method to yield accurate results, stages have to be
 placed at appropriate  values of $\lambda$ such that successive
 distributions $p(N_{\rm o}|\lambda_i)$ and $p(N_{\rm
   o}|\lambda_{i+1})$ overlap significantly. This is essentially the
 same criteria for choosing the set of intermediates
 $\{\lambda^{({\rm m})}\}$ that is required to yield a reasonable acceptance
 rate between all stages in method I (Sec.~\ref{sec:methodI}). Indeed
 comparing with Eq.~\ref{eq:accprob1}, one sees that
 Eq.~\ref{eq:zwanzig1} provides a measure of the acceptance rate for
 transitions between neighbouring stages as explicitly implemented in
 method I. Accordingly it serves as a basis for thinning out,
 appropriately, the trial set of $1000$ stages as described in
 Sec.~\ref{sec:methodIremarks}. The resulting set $\{\lambda^{({\rm m})}\}$ is
 then equally applicable to methods I and II.  We emphasize that for either
 method there is no need to recalculate the set $\{\lambda^{({\rm m})}\}$
 for each $r_{\rm bb}$ of interest; determining a set for one value of
 $r_{\rm bb}$ suffices for all values provided the depletion potential
 does not vary by more than a few $k_BT$.  We also remark in
 passing that while method II bears some resemblance to thermodynamic
 integration schemes \protect\cite{Frenkelsmit2002}, the estimates of
 the free energy differences are in principle exact- no numerical
 quadrature is involved.

\begin{figure}[h]
   \includegraphics[type=pdf,ext=.pdf,read=.pdf,width=0.95\columnwidth,clip=true]{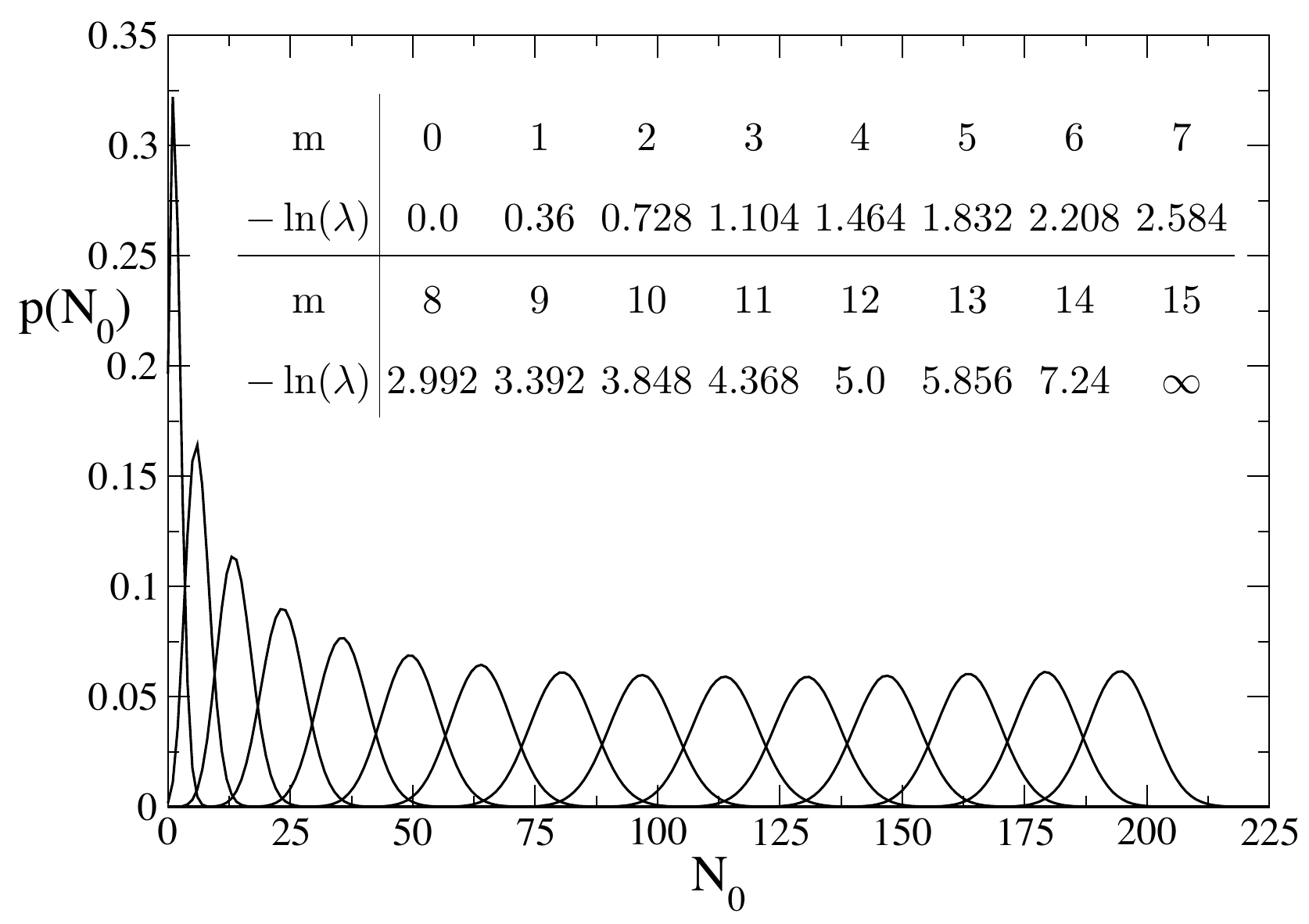}
 \caption{The measured form of the overlapping distributions
   $p(N_{\rm o}|\lambda^{({\rm m})})$ employed to measure the shell
   insertion probability for $r_{\rm bb}=1.06, \eta_{\rm s}^r=0.32,q=0.1$ via method II.
   From right to left the distributions correspond to increasing ${\rm m}$ from ${\rm m}=0$ to
   ${\rm m}=14$. The corresponding values of $\ln \lambda$ are shown in the key. The distribution for ${\rm m}=15$ is not depicted as it encompasses
   only the $N_o=0$ state.}
   \label{fig:pN0}
 \end{figure}

Fig.~\ref{fig:pN0} shows our measurements of the set of $M=16$
individual distributions $p(N_{\rm o}|\lambda^{({\rm m})})$ for
$\eta_{\rm s}^r=0.32, q=0.1$ that yield an estimate of the insertion
probability via application of Eqs.~\ref{eq:zwanzig2} and \ref{eq:whtdiff}. The set $\{\lambda^{({\rm m})}\}$ is the same as that
  used in method I and is listed in the key.

 The chief merit of the multiple overlapping histogram approach
 compared to the expanded ensemble approach (method I) is its
 simplicity: no weights need to be calculated before one can start to
 accumulate data. Its main disadvantage compared to method I, is the
 need to perform $M$ independent simulations and synthesize the
 results in a pairwise fashion. However, this drawback is somewhat mitigated 
by the fact that the independence of the simulations for
 each $\lambda^{({\rm m})}$ renders them trivially parallel. Accordingly,
 one can farm out the calculations for each to a separate processor on
 a multiprocessor computer. Similarly the estimates of the insertion
 probability at the various values of $r_{\rm bb}$ that are needed to construct
 the full depletion potential are also independent, and can therefore be
 accumulated in parallel.

 \subsection{Method III: Umbrella sampling}

 This approach, which has some commonality with the umbrella sampling approach 
 of Ding and Valleau \cite{Ding:1993ys}, is conceptually simpler than the
 previous two in that it dispenses with staged intermediates.

 \subsubsection{Description}

 The algorithm considers an imaginary shell of diameter $\sigma_b$
 centered on $x=r_{\rm bb}$. The instantaneous number of
 small particles, $N_{\rm o}$, that overlap this notional shell fluctuates with time, and hence
 one can measure its distribution $p(N_{\rm o}|r_{\rm bb})$ as a histogram. Typically $N_{\rm o}$ will
 be large, but we can performs biased (``Umbrella'') sampling
 with respect to insertion and deletion of the small particles in order
 to accurately measure the probability of states having $N_{\rm o}=0$. A little thought
 shows that this probability is just the shell insertion probability
 required for Eq.~\ref{eq:effpot_ins}.

Operationally, transfers of small particles are performed according to the biased acceptance probabilities:

 \bea
   p_a(N_{\rm s} \!\to \!N_{\rm s}\!+\!1)&=&{\rm min}\!\left(1,\frac{V}{N_{\rm s}\!+\!1} e^{\mu + W^+}\right)\;,\nonumber\\
   p_a(N_{\rm s}\! \to\! N_{\rm s}\!-\!1)&=&{\rm min}\!\left(1,\frac{N_{\rm s}}{V} e^{-\mu+ W^-}\right)\:.\nonumber\\
   \label{eq:gceacc}
 \eea
These are the standard criteria for the grand canonical ensemble
\cite{Frenkelsmit2002}, modified by a weight factor $W^\pm$ 
that is non-zero if the proposed insertion or deletion of a small
particle leads to a change in the number of overlaps $N_{\rm o}$.
Specifically 

\bea
W^+ &=& w\left(N_{\rm o}(\{{\bf r}\}^{N_s+1})\right)-w\left(N_{\rm o}(\{{\bf r}\}^{N_{\rm s}})\right),\nonumber\\
W^- &=& w\left(N_{\rm o}(\{{\bf r}\}^{N_s-1})\right)-w\left(N_{\rm o}(\{{\bf r}\}^{N_{\rm s}})\right).\nonumber\\
\label{eq:weights}
\eea
Here $N_{\rm o}(\{{\bf r}\}^{N_{\rm s}})$ is the number of overlap arising
from the set of position vectors $\{{\bf r}\}^{N_{\rm s}}={\bf r}_1,{\bf r}_2\ldots{\bf
r}_{N_{\rm s}}$ of $N_{\rm s}$ small particles, while $w(N_{\rm o})$ is a weight
function defined on the number of overlaps.  These weights allow a
single simulation run to sample not just the values of $N_{\rm o}$
that are typical for a given $\eta_{\rm s}^r$, but also the entire
range down to $N_{\rm o}=0$. Accordingly one can measure a histogram
of the weighted probabilities $\tilde{H}(N_{\rm o}|r_{\rm bb})$, from
which the Boltzmann histogram is obtained by unfolding the weights:

\be
H(N_{\rm o}|r_{\rm bb})=\tilde{H}(N_{\rm o}|r_{\rm bb})e^{w(N_{\rm o})}\:.
\ee
After normalization, this yields the probability distribution $p(N_{\rm
  o}|r_{\rm bb})$, from which the insertion probability is read off as
$p(N_{\rm o}=0|r_{\rm bb})$. The depletion potential (up to a
constant) follows via eq.~\ref{eq:effpot_ins}.  Repeating for a
sequence of values of $r_{\rm bb}$ allows one to build up the entire
depletion potential.

 \subsubsection{Remarks and results}

\begin{figure}[h]

   \includegraphics[type=pdf,ext=.pdf,read=.pdf,width=1.00\columnwidth,clip=true]{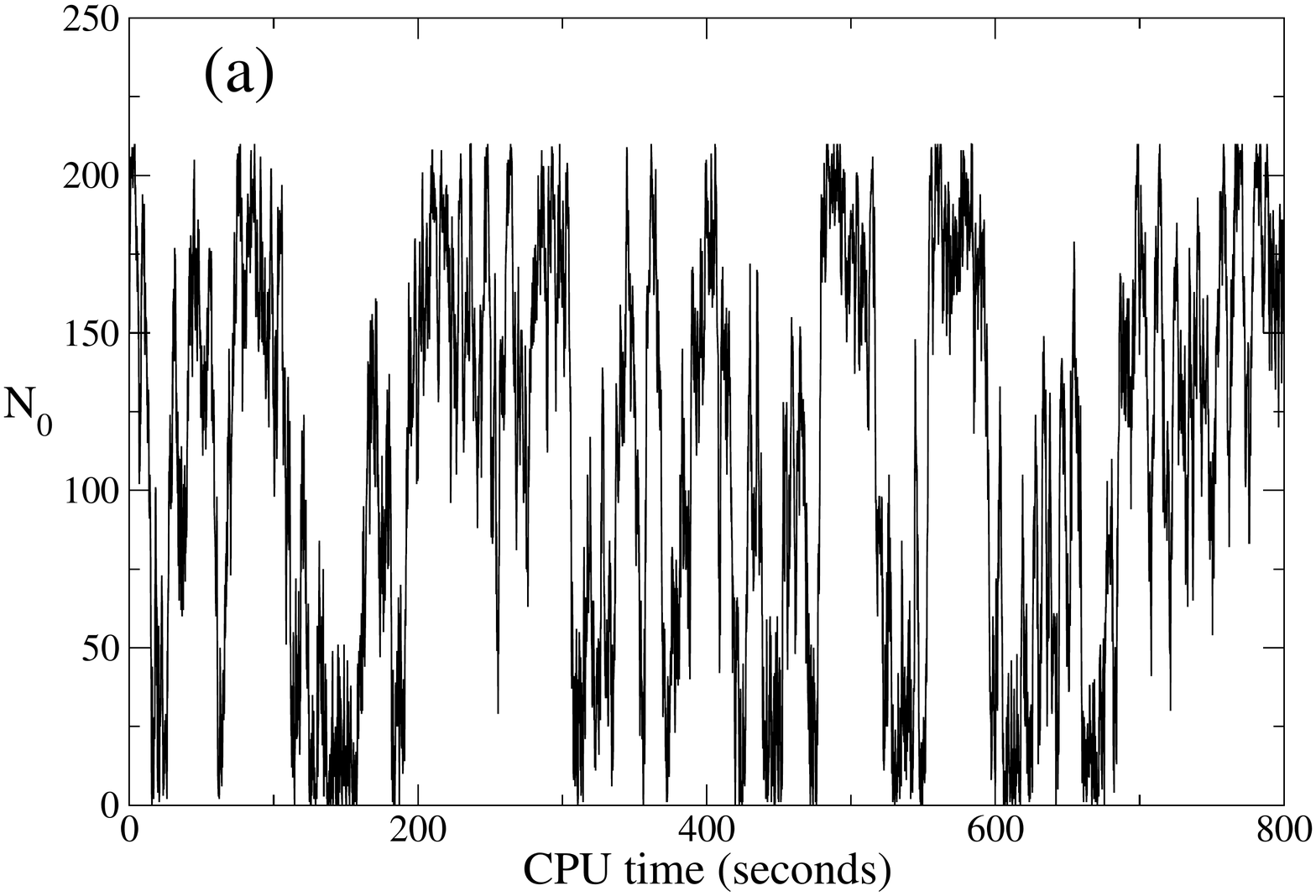}
   \includegraphics[type=pdf,ext=.pdf,read=.pdf,width=1.00\columnwidth,clip=true]{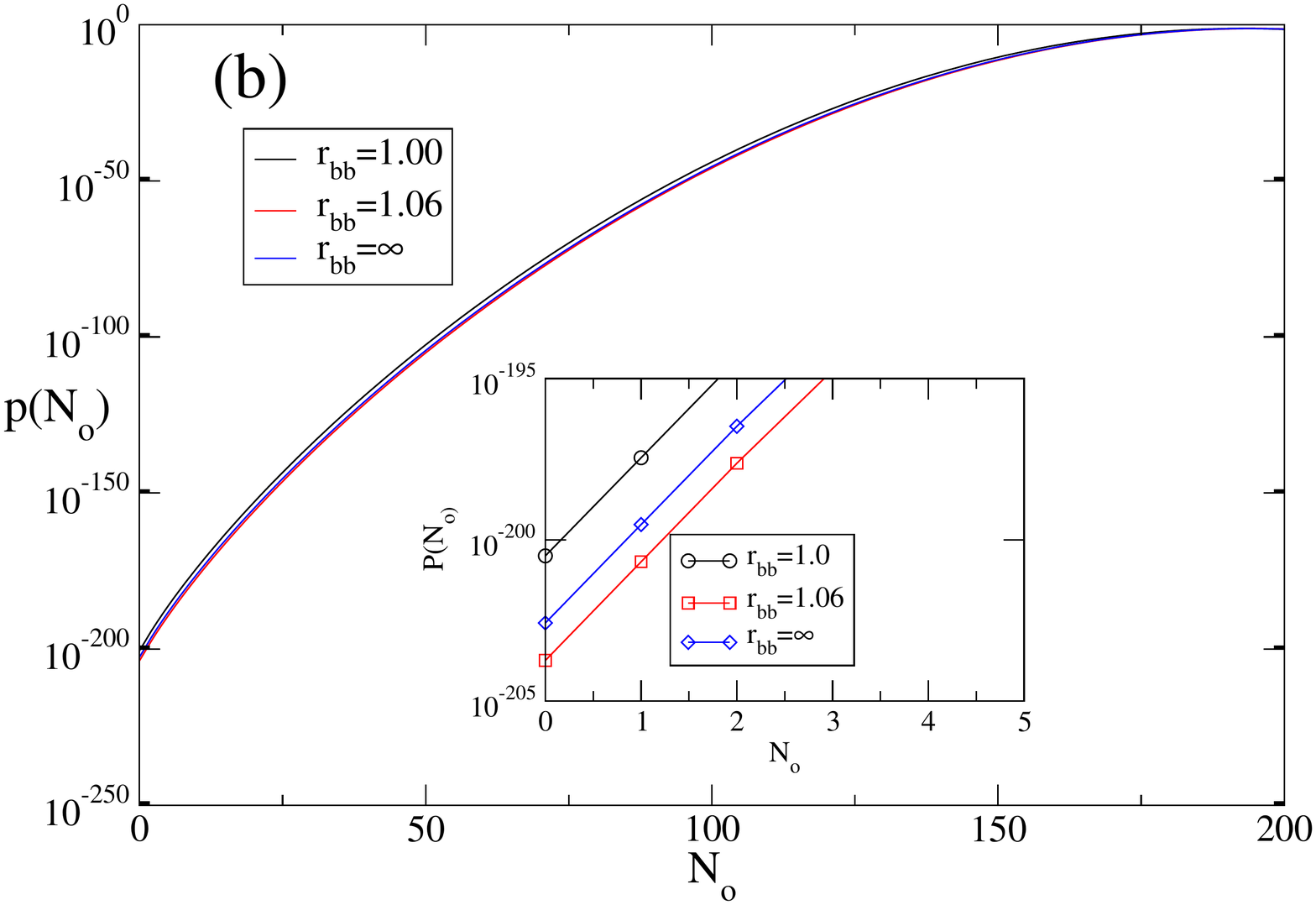}

    \caption{{\bf (a)} $N_{\rm o}(t)$ {\em vs} CPU time on a $2$ GHz
      processor at $r_{\rm bb}=1.06\sigma_{\rm b}, \eta_s=0.32,q=0.1$ as
      obtained from method III. The figure gives an impression of the typical
      timescale required to sample the range from $N_o=\bar{N_o}$ to
      $N_o=0$, but constitutes only a small portion of the full run
      which comprised $35$ CPU hours. {\bf (b)} The form of the
      overlap probability distribution $p(N_{\rm o}|r_{\rm bb})$ at
      three values of $r_{\rm bb}$. Differences in $p(N_{\rm
        o}=0|r_{\rm bb})$ as a function of $r_{\rm bb}$ (inset) yield
      the depletion potential as described in the text.}
   \label{fig:methodIII}
 \end{figure}

As with method I, an appropriate set of weights is required for this
method to operate effectively and again these can be readily
determined using the TMMC method (Appendix~\ref{sec:tmmc}).
Fig.~\ref{fig:methodIII} shows a time series of the sampled values of
$N_{\rm o}$ that results once the weights are in place. Owing to the
biasing, the system samples smoothly the entire range from the most
probable number of overlaps $N_{\rm  o}\approx 200$, right down to $N_{\rm
  o}=0$. The resulting form for $p(N_{\rm o}|r_{\rm bb})$, obtained by
unfolding the effects of the weights and normalizing the resulting
histogram is shown in Fig.~\ref{fig:methodIII}(b). From this one
simply reads off the shell insertion probability as $p(N_{\rm
  o}=0|r_{\rm bb})$.

 The chief merit of method III compared to methods I and II is that it
 is parameter free: there are no staged intermediates and therefore
 the associated inconvenience and startup costs of determining their
 number and appropriate placement are obviated. Nevertheless the
 computational cost of calculating weights represents a significant
 overhead as will be discussed in Sec.~\ref{sec:discuss}. We note that method III
 is parallelisable, but only with respect to the separate measurements
 at various $r_{\rm bb}$ needed to build up the depletion potential.

 \section{Direct sampling route: Implementations}
 \label{sec:cluster}

 We now turn to consider two schemes that accumulate the depletion potential by
 focusing on the difference in effective potential as one varies
 $r_{\rm bb}$. They both rely on collective (cluster) updates of big and
 small particles. One is based on the cluster algorithm of Dress and
 Krauth \cite{Dress1995}, the other is a bespoke constrained cluster
 algorithm.

 \subsection{Method IV: Geometrical Cluster Algorithm}
 \label{sec:gca}

 An efficient cluster algorithm capable of dealing with hard spheres
 mixtures was introduced by Dress and Krauth in 1995 \cite{Dress1995}.
 It was subsequently generalized to arbitrary interaction potentials
 by Liu and Luijten \cite{Liu2004,Liu2005} who dubbed their method the
 Geometrical Cluster Algorithm (GCA). A restricted Gibbs ensemble
 version of the GCA suitable for studying phase transitions was also
 subsequently developed \cite{Liu2006,Ashton2010,Ashton:2010ys}. Here
 we describe the GCA for a general system of hard spheres in the
 canonical ensemble, before specializing to the case of a size
 asymmetrical binary mixture.

 \subsubsection{Description}

 The particles comprising the system are assumed to be contained in a
 periodically replicated cubic simulation box of volume~$V$.  The
 configuration space of these particles is explored via cluster updates, in
 which a subset of the particles known as the ``cluster'' is displaced
 via a point reflection operation in a randomly chosen pivot point. The
 cluster generally comprises both big and small particles and by virtue
 of the symmetry of the point reflection, members of the cluster retain
 their relative positions under the cluster move. Importantly, cluster
 moves are rejection-free even for arbitrary interparticle interactions
 \cite{Liu2005}. This is because the manner in which a cluster is built
 ensures that the new configuration is automatically Boltzmann
 distributed.

 For hard spheres (there is no advantage in using shells in this
 context), the cluster is constructed as follows: one of the particles
 is chosen at random to be the seed particle of the cluster.  This
 particle is point-reflected with respect to the pivot from its
 original position to a new position.  However, in its new position,
 the seed particle may overlap with other particles. The identities of
 all such overlapping particles are recorded in a list or
 ``stack''. One then takes the top-most particle off the stack, and
 reflects its position with respect to the pivot. Any particles which
 overlap with this particle at its destination site are then added to
 the bottom of the stack. This process is repeated iteratively until
 the stack is empty and there are no more overlaps.

 Note that cluster updates only displace particles, they do not
 allow their number to fluctuate. Accordingly, in order to treat the small particles grand
 canonically, we also perform insertions and deletions of small
 particles with a chemical potential corresponding to the prescribed
 $\eta_{\rm s}^r$, as outlined in Sec.~\ref{sec:system},

 The effective potential $W(r)$ between two big particles is defined
 in terms of the radial distribution function $g(r_{\rm bb})$,  measured in the limit of infinite
 dilution:

 \be
 W(r)=-\lim_{\rho_b\to 0}\ln[g(r_{\rm bb})] \:,
\label{eq:wfromg}
 \ee
 for $r_{\rm bb}>\sigma_b$. In our simulation studies this limit is approximated by
 placing a single pair of big hard spheres in the simulation box.  A
 finite-size estimate to $g(r_{\rm bb})$, which we shall denote $g_L(r_{\rm bb})$, is then
 obtained by fixing the first of these particles at the origin and
 measuring (in the form of a histogram) the probability $p(r_{\rm bb})$ of finding the
 second big particle in a shell of radius $r_{\rm bb}\to r_{\rm bb}+dr$. Then

 \be
 g_L(r_{\rm bb})=\frac{p(r_{\rm bb})}{p_{\rm ig}(r_{\rm bb})}\:,
 \label{eq:g_r}
 \ee
 where the normalization relates to the probability of finding an ideal gas
 particle at this radius:

 \be
 p_{\rm ig}(r_{\rm bb})=\frac{4\pi r^2}{V}\:.
 \label{eq:Pig}
 \ee

 To effect the measurement of $g_L(r_{\rm bb})$, we modify the GCA slightly as follows: we
 choose one big particle to be the seed particle, which we place
 randomly within a shell $\sigma_{\rm b} < r_{\rm bb} < L/2$, centered on the second
 big particle, with $L$ the linear box dimension. The location of the
 pivot is then inferred from the old and new positions of the seed
 particle. Thereafter clusters are built in the standard way. This
 strategy ensures that we efficiently sample separations of the big
 particles that lie in the range $\sigma_{\rm b} < r_{\rm bb} < L/2$ for which
 $g(r_{\rm bb})$ can sensibly be defined for hard spheres in a cubic box. 

 \subsubsection{Remarks and results}

 For the systems of interest in this work, we find that the GCA is
 efficient for reservoir packing fractions $\eta_{\rm s}^r \le 0.2$. Above this
 value, practically all the particles join the cluster,
 which merely results in a trivial point reflection of the entire
 system. Indeed the efficiency drop is so precipitous that
 $\eta_{\rm s}^r=0.2$ is the absolute upper bound on the volume fraction of small
 particles that can usefully be studied with this algorithm. For single
 component fluids this problem can be ameliorated by biasing the choice
 of pivot position to be close to the position of the seed particle
 \cite{Liu2005}. Doing so has been reported to extend the operating
 limit to $\eta_{\rm s}^r\simeq 0.34$.  However, for the case of highly
 asymmetrical mixtures we find that this strategy does not
 significantly decrease the number of particles in the cluster because
 as soon as a big particle joins the cluster and is point reflected it
 causes many overlaps with small particles.

\begin{figure}[h]
  \includegraphics[type=pdf,ext=.pdf,read=.pdf,width=0.95\columnwidth,clip=true]{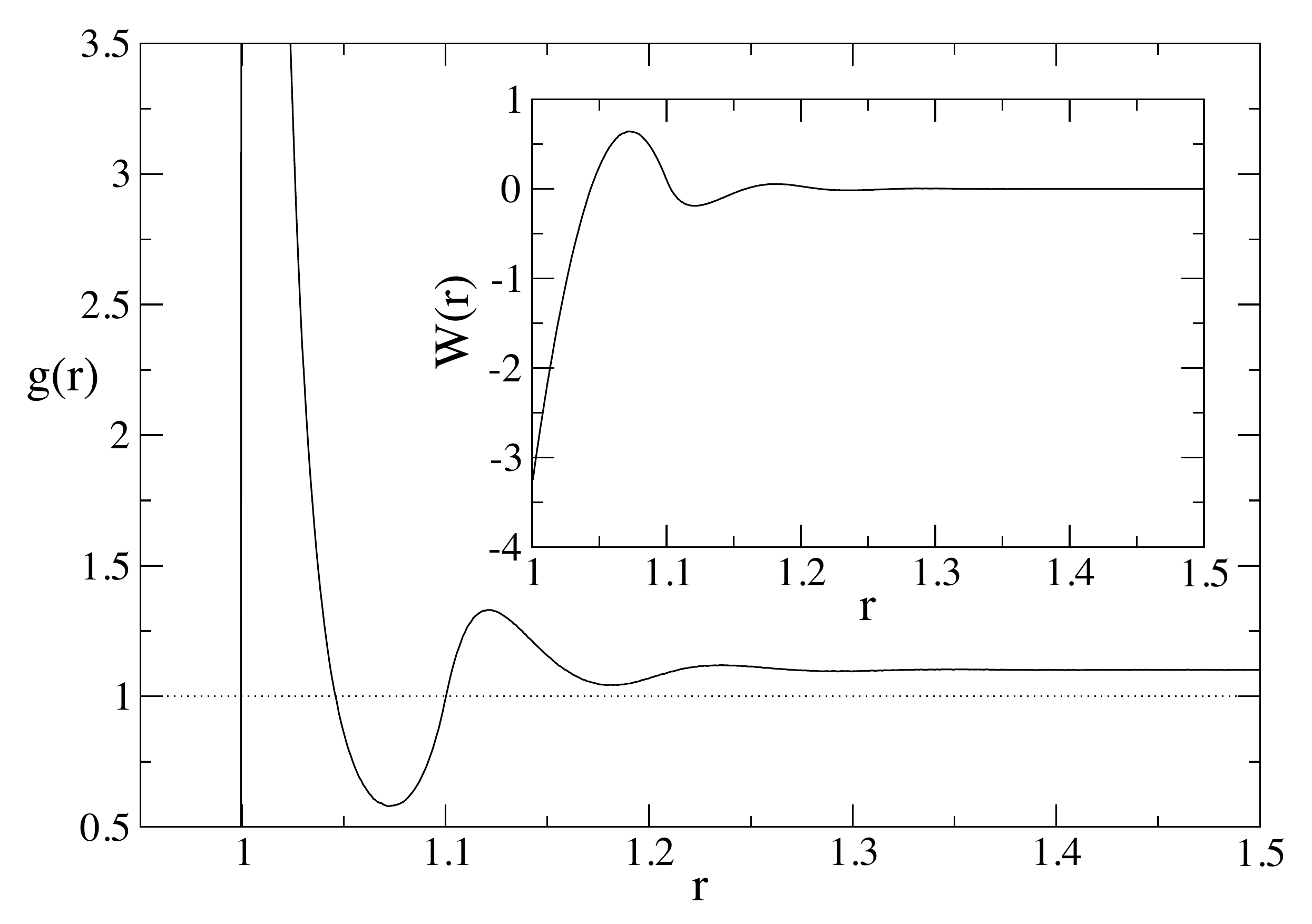}
    \caption{The measured form of $g_L(r_{\rm bb})$ corresponding to
      $\eta_{\rm s}^r=0.2, q=0.1$, obtained via method IV for a simulation box of
      dimensions $V=(3\sigma_b)^3$. The limiting value differs from
      unity due to the finite-size effects described in Ashton {\em et al}\;\protect{\cite{Ashton:2011kx}}. The inset shows
       the depletion potential $W(r)$ obtained by implementing
      the finite-size correction described in the text to $g_L(r_{\rm bb})$ and applying Eq.~\ref{eq:wfromg}.
}
   \label{fig:g_r}
 \end{figure}

 Fig.~\ref{fig:g_r}(a) shows the measured form of $g_L(r_{\rm bb})$ for
 $\eta_{\rm s}^r=0.2, q=0.1$ obtained using a cubic simulation box of volume
 $V=(3\sigma_b)^3$. For this measurement to provide an estimate of
 $W(r_{\rm bb})$, it first has to be corrected for finite-size effects,
 manifest in the failure of the function to approach unity at large
 $r_{\rm bb}$. This is done (as has also been described
 elsewhere\cite{Ashton:2011kx}) by measuring the cumulative integral

 \be
 G(R)=\int_0^{R}g_L(r)dr \:.
 \ee
 This integral tends towards a smooth linear form quite rapidly as the
 upper limit $R$ increases.  The measured limiting gradient, $\xi$, of
 $G(R)$ provides the requisite correction factor according to $g(r)=
 \xi^{-1}g_L (r_{\rm bb})$. Following Eq.~\ref{eq:wfromg}, the negative of the
 logarithm of $g(r)$ then yields an estimate for the effective
 potential $W(r_{\rm bb})$, which is shown in Fig.~\ref{fig:g_r}(b).

 The most attractive feature of the GCA for determining depletion
 potentials is that it allows direct sampling of the quantity of
 interest without the need for multiple simulations or biased
 sampling.  Its principal drawback is that the method becomes unusable
 for $\eta_{\rm s}^r\gtrsim 0.2$, which limits its applicability. It
 is therefore of interest to consider whether one can formulate an
 algorithm that exploits the efficiency of collective updates, but
 operates at higher values of $\eta_{\rm s}^r$. The method described in the
 following subsection achieves this, albeit at the expense of
 introducing biased sampling.

 \subsection{Method V: Constrained cluster algorithm}

In common with the GCA, this method collectively moves a big hard
sphere and a number of small ones via a self inverse
operation. However, in contrast to the GCA it is a constrained scheme in
the sense that it measures the free energy differences between two
neighbouring discrete values of $r_{\rm bb}$. 

 \begin{figure}[h]
 \includegraphics[type=pdf,ext=.pdf,read=.pdf,width=0.95\columnwidth,clip=true]{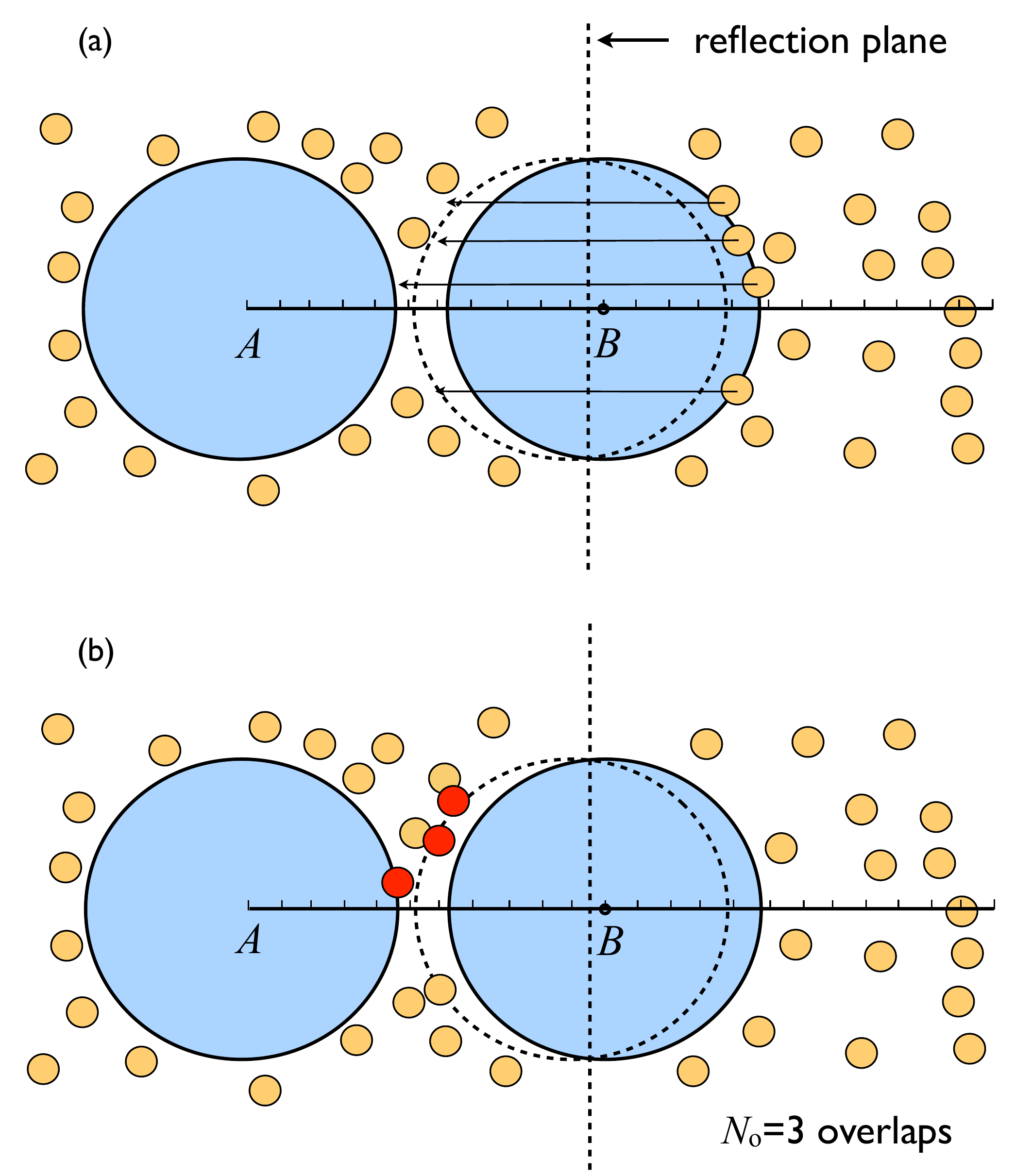}

 \caption{Schematic illustration of the constrained cluster
   update of method V. {\bf (a)} The big particle $B$ at $x_i$ undergoes a plane
   reflection to $x_{i+1}$ and thereby overlaps with $n$ small
   particles which are themselves subsequently reflected in the
   plane. {\bf (b)} In their new position, the $n$ small particles
   overlaps with $N_{\rm o}$ other particles (one of which may be the
   big particle $A$ at the origin). The number of such secondary overlaps
   $N_{\rm o}$ is the primary observable.}

 \label{fig:method5}
 \end{figure}


 \subsubsection{Description}

The operation of the method is shown schematically in
Fig.~\ref{fig:method5}.  One big hard sphere (particle $A$) is fixed
at the origin. The other (particle $B$) can occupy discrete values of
$r_{\rm bb}$ set out along a one-dimensional radial grid which we take
to be the $x$-axis. Let us label the grid points by the index $i$, and
consider the situation when the big particle is stationed at $r_{\rm
  bb}=x_i$. We then estimate the free energy difference between 
grid points $i$ and $i+1$ in the following manner.

With particle $B$ stationed at grid point $i$, we equilibrate the
small particles via transfers with the reservoir. For some equilibrium
configuration of the small particles we then consider (but do not
implement) a trial move to take particle $B$ from grid point $i$ to
grid point $i+1$ as follows:

 \begin{enumerate}

 \item Reflect the center of particle $B$ in the plane normal to the $x$
 axis which cuts the $x$ axis at $x= (x_i+x_{i+1})/2$. This takes particle
 $B$ from grid point $i$ to grid point $i+1$ as shown in fig.~\ref{fig:method5}(a).

 \item Under this move, particle $B$ will overlap with a number
   $n$, say, of small particles. We then imagine reflecting these
   $n$ small particles in the same reflection plane. This switches
   them into the space left by particle $B$, see fig.~\ref{fig:method5}(a).

 \item After undergoing this reflection, some of the $n$ small particles will overlap
   with other small particles or with the big particle $A$, as shown
   in fig.~\ref{fig:method5}(b). The number of such `secondary' overlaps is the
   observable $N_{\rm o}$ for the current configuration of small
   particles.

 \end{enumerate}

 One then samples the fluctuations in $N_{\rm o}$ with respect to the
 ensemble of small particle configurations and accumulates its
 probability distribution $p(N_{\rm o})$ as a histogram. Similarly to
 methods I-III, it is beneficial to preferentially implement transfers
 of small particles in a shell region around big particle $B$;
 this concentrates the computational effort on those regions which
 contribute most to the measurement. The sampling of the small
 particle configurations is biased so as to enhance the occurrence of
 values of $N_{\rm o}$ down to $N_{\rm o}$=0. This is achieved by
 defining a weight function $w(N_{\rm o})$ which is incorporated in
 the GCE acceptance probabilities Eq.~\ref{eq:gceacc}, in exactly the
 same manner as described for method III. An appropriate weight
 function can be found automatically using the TMMC method described
 in the appendix.

Together these measures enable an efficient and accurate
estimate for the probability that the trial collective move leads to
$N_{\rm o}=0$, ie. a valid hard sphere configuration.  Let us denote
this probability $p_i^{\scriptstyle +}(0)$ because we have measured it
with particle $B$ moving from grid point $i$ to $i+1$. Similarly we
can measure the probability $p_{i+1}^{\scriptstyle -}(0)$ that a move
from $i+1\to i$ leads to zero overlaps. Then the measured ratio
$p_i^{\scriptstyle +}(0)/p_{i+1}^{\scriptstyle -}(0)$ provides the
difference in the depletion potential between grid points $i$ and
$i+1$ via an an expression akin to Bennett's acceptance ratio formula
\cite{Bennett:1976uq}:

 \be
W(x_{i+1})- W(x_i)=\ln\frac{p_{i+1}^{\scriptstyle -}(0)}{p_i^{\scriptstyle +}(0)}\:.
 \label{eq:DeltaW}
 \ee
From measurements of the difference in the depletion
potential between all neighbouring pairs of grid points, one extracts
the depletion potential itself simply by summing, commencing at a
value of $r_{\rm bb}$ sufficiently large that $W(r_{\rm bb})$
can be considered to have decayed to zero.

 \subsubsection{Remarks and results}

Compared to the GCA (method IV), the principal asset of method V is
that it permits study of considerably larger volume fractions of the
small particles.  This is because the number of particles involved in
the collective move is not allowed to grow indefinitely. Instead
cluster growth is truncated after one iteration and biased sampling
used to obtain the information required to estimate the depletion
potential. We note that a constrained cluster algorithm suitable for
estimating depletion potentials has previously been described by
Malherbe and Krauth \cite{Malherbe:2007vn}, however it does not
truncate cluster growth and therefore is limited to much lower values
of $\eta_{\rm s}^r$ than the present approach.

\begin{figure}[h]
 \includegraphics[type=pdf,ext=.pdf,read=.pdf,width=0.95\columnwidth,clip=true]{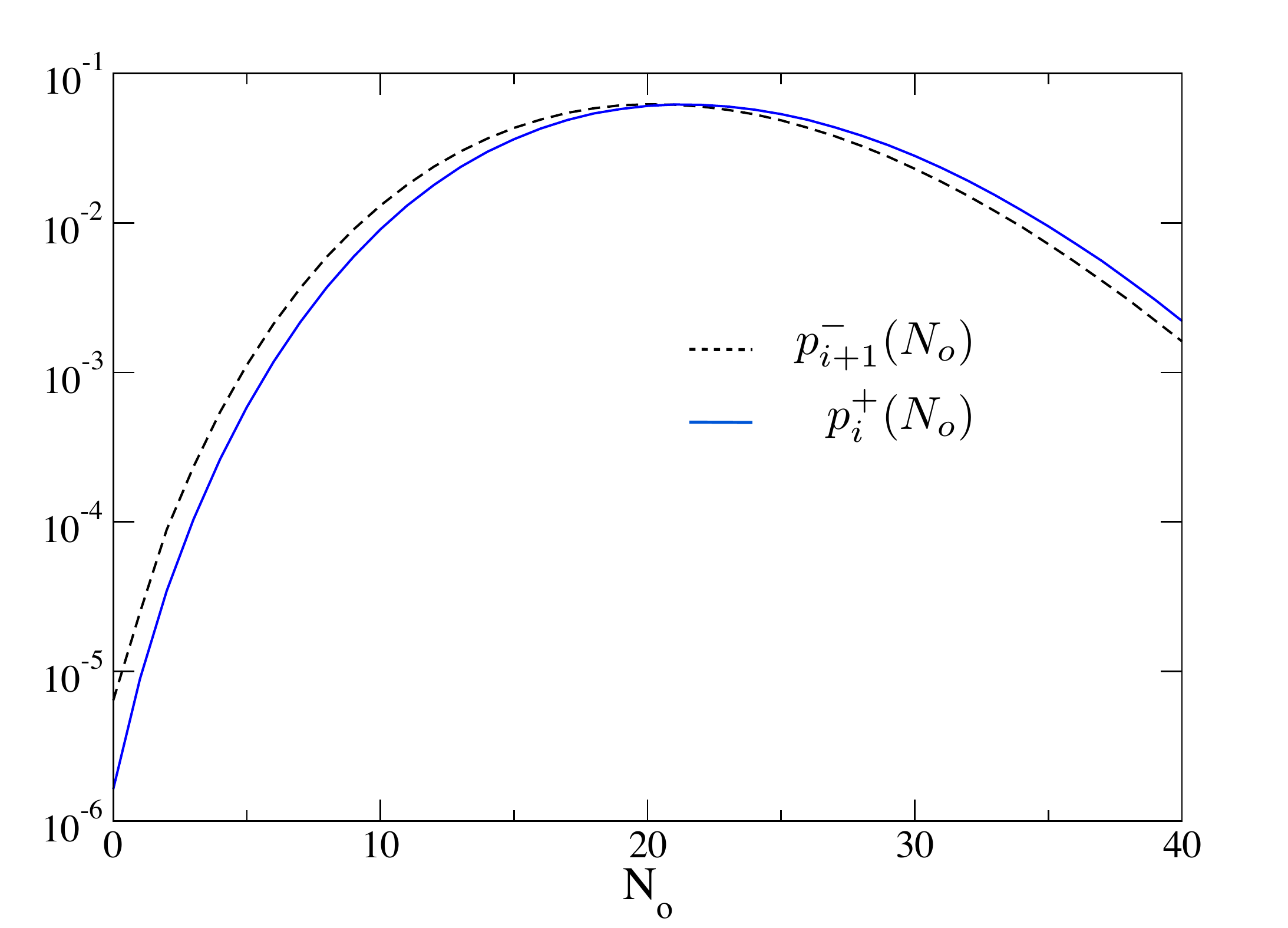}
 \caption{Estimates of $p_i^{\scriptstyle +}(N_{\rm o})$ and $p_{i+1}^{\scriptstyle -}(N_{\rm o})$ for $i=0$, corresponding to
  contact of the big particles (i.e. $r_{\rm bb}=1.0\sigma_{\rm b}$) as obtained using method V. The grid point separation is $x_{i+1}-x_i=0.05$. The
   ratio of the values of these functions for $N_{\rm o}=0$  provides an estimate of the
   difference in the effective potential between the grid points via Eq.~\protect\ref{eq:DeltaW}.}
 \label{fig:method5dists}
 \end{figure}

In common with the gradual insertion methods I and III, the
constrained cluster algorithm requires (in general) knowledge of a set of weights
for its operation.  However, because the method focuses on free
energy differences, the typical number of overlaps ${\overline N_{\rm
    o}}$ is generally far fewer than encountered in methods I and III,
and hence the degree of weighting required to reach $N_{\rm o}=0$ is
much less. For example, for $\eta_{\rm s}^r=0.32$, and a grid point
separation of $x_{i+1}-x_i=0.05$ we find ${\overline N_{\rm o}}\approx
20$ (see Fig.~\ref{fig:method5dists}) which is to be compared with the
$\approx 200$ overlaps that occur for shell insertion in methods I and
III. Thus weight calculation is relatively quick and easy for
method V, and indeed we find that if we reduce the small particle volume
fraction to $\eta_{\rm s}^r\lesssim 0.2$, then no weights are
required at all since the system samples the $N_{\rm o}=0$ state
sufficiently often without the aid of biasing.  Even in cases where
weighting is required, it is in general not necessary to calculate
weights for every grid point; to the extent that the effective
potential does not vary strongly between grid points, weights found
for one grid point will suffice for all other grid
points. 

Although it is perhaps reminiscent of methods that obtain the
depletion potential by integrating the measured force in an MD setting
\cite{Biben1996,Dickman1997,Goetzelmann1999,Herring07}, method V
provides {\em exact} differences in the depletion potential i.e. no
quadrature is required. However, one downside of the need to sum free
energy differences to obtain the depletion potential is that
cumulative errors arise. The error grows with the number of
differences summed and can potentially lead to an estimate for
$W(r_{\rm bb})$ that whilst appearing quite smooth, nevertheless
deviates significantly from the exact form. Since we commence summing
the free energy differences at large $r$, where the potential can
be assumed to be essentially zero, this implies that the largest
errors occur near contact. To be more precise, for $j=1\cdots N$ free
energy differences, the variance in the sum is simply the sum of the
variances of the individual (uncorrelated) estimates ie
$\sigma_N^2=\sum_{j=1}^N\sigma^2_j$.  If each individual measurements
receives an equal computational expenditure then to a good
approximation the cumulative error after summing $N$ differences is
simply $\sigma_N=\sqrt{N}\sigma$. This growth in the uncertainty 
in the estimate of $W(r)$ as $r$ decreases, contrasts with the gradual
insertion methods where every point in the estimate of $W(r_{\rm bb})$
is independent.

Finally in this subsection we remark that in common with methods
I-III, method V is parallelisable with respect to calculations along
the grid: one can simply set up independent copies of the simulated
system each of which calculates $W(x_{i+1})- W(x_i)$ for a different
grid point $i$.

\section{Discussion}
\label{sec:discuss}

In the preceding two sections we have described five distinct
methods for determining depletion potentials in highly size
asymmetrical hard sphere mixtures. We now turn to a discussion of
their relative merits.

\begin{figure}[h]
\includegraphics[type=pdf,ext=.pdf,read=.pdf,width=0.95\columnwidth,clip=true]{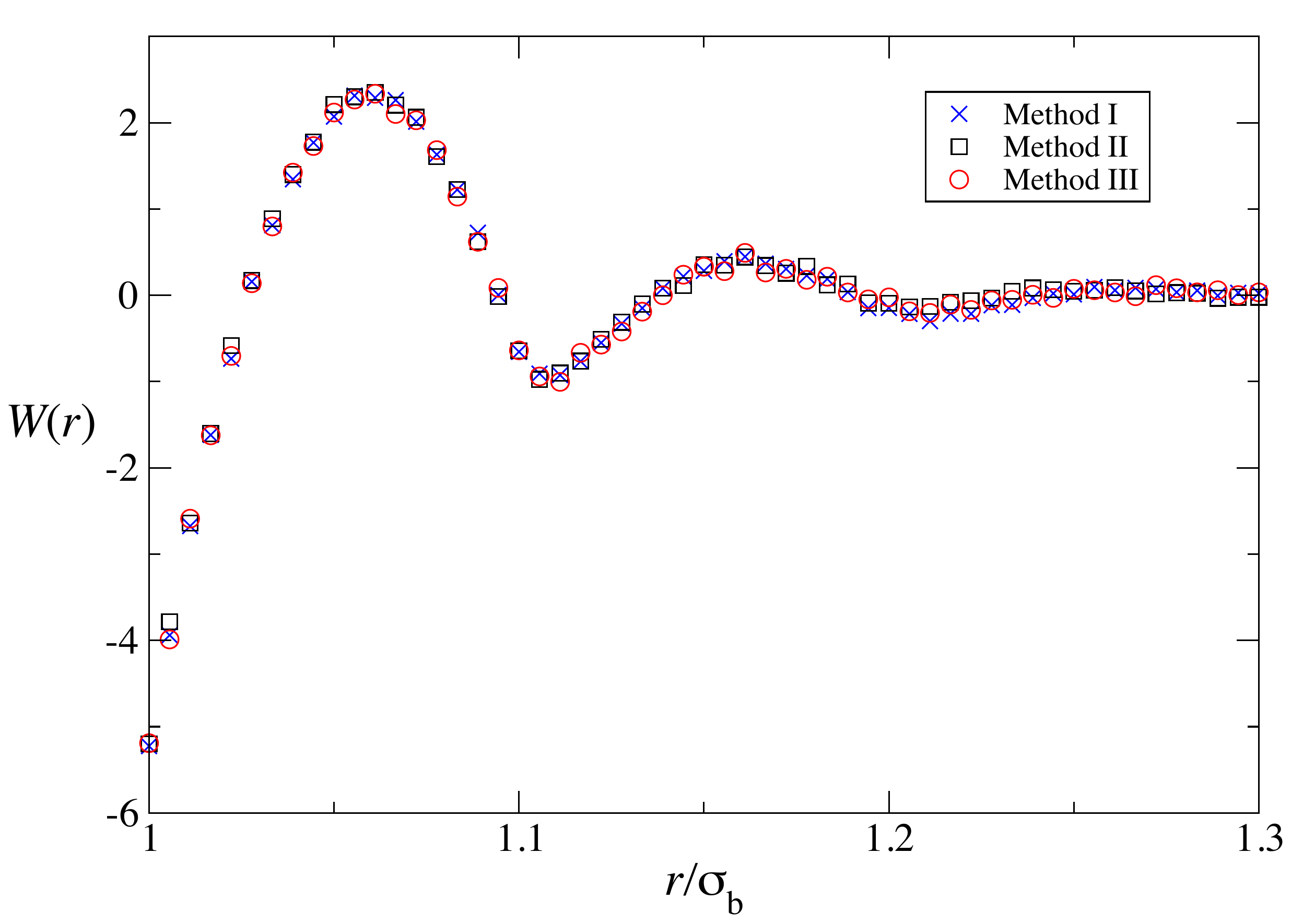}
\includegraphics[type=pdf,ext=.pdf,read=.pdf,width=0.95\columnwidth,clip=true]{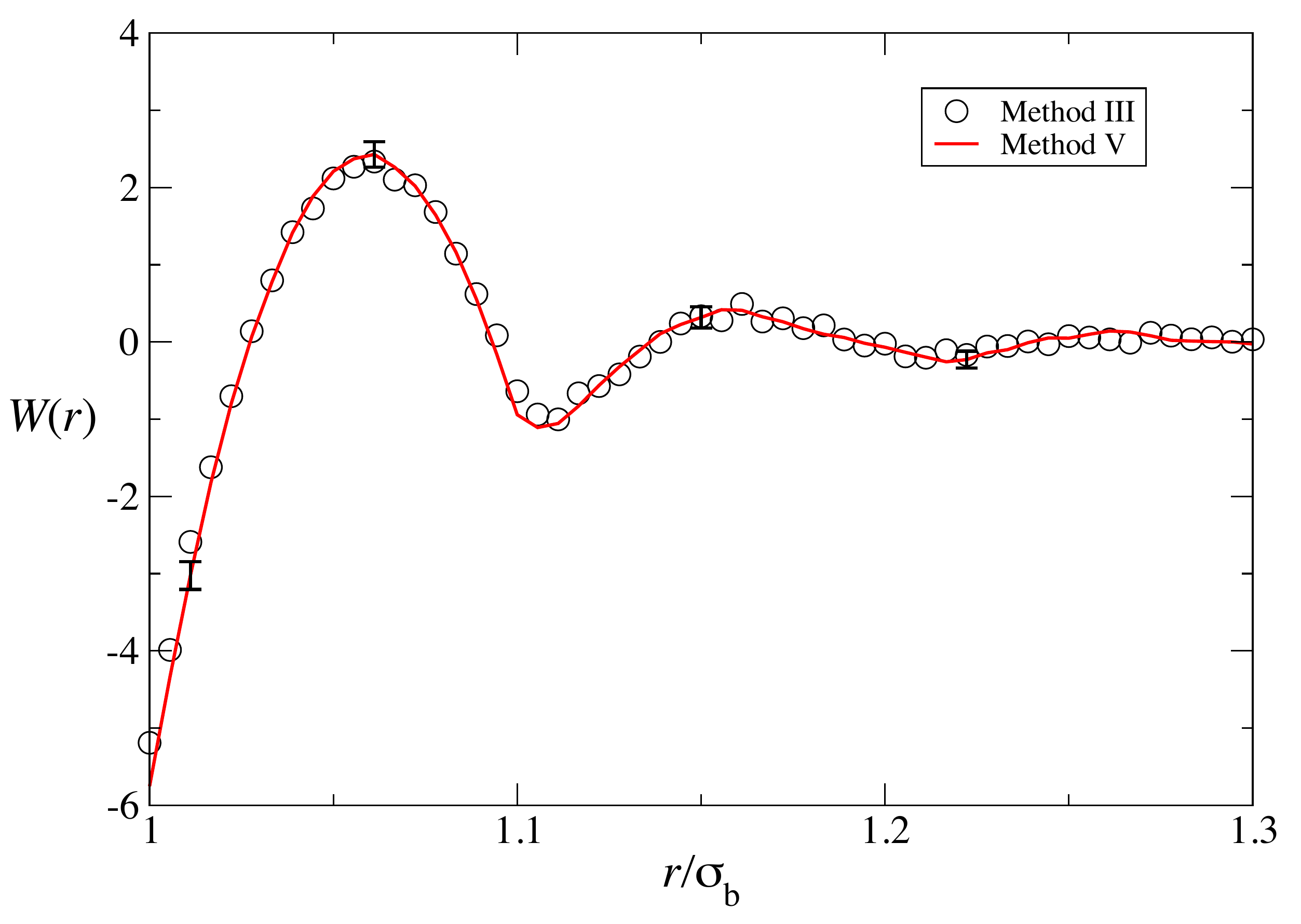}

 \caption{Estimates of the form of $W(r_{\rm bb})$ for $q=0.1,
   \eta_s^r=0.32$.  {\bf (a)} Comparison of the results of methods I-III.
       {\bf (b)} Comparison of the results of methods I and III. All data
       corresponds to a computational investment of $35$ CPU hours per
       points on a $2$ GHz processor. For methods I-III statistical
       errors are comparable with symbol sizes. For method V,
       the statistical error accumulates in going from large to small $r$ as indicated by the representative error bars.}
 \label{fig:overall}
 \end{figure}

Let is begin by comparing the gradual insertion methods I-III amongst
themselves. In terms of their relative efficiency, we find that once
prepared so that sampling can commence, each of the methods I-III take
a similar amount of CPU time to achieve a given statistical accuracy
for $W(r_{\rm bb})$. This is shown in Fig.~\ref{fig:overall}(a) which
displays the form of the depletion potential at $\eta_s=0.32$ for
$q=0.1$ as obtained from methods I-III. The same amount of CPU time
($35$ hours per point on a $2$ GHz processor) was invested in each
method, and the curves are comparable with respect to smoothness.
This finding is perhaps not surprising since in one way or another
they all seek to bias small particles out of the way so that one can
calculate the insertion probability of a big hard shell.

However, when one takes into account computational startup costs,
significant differences arise in the overall efficiencies of methods
I-III.  Method I is the most cumbersome of the three in this respect
since it requires both the choice of a suitable set of staged
intermediates and knowledge of a set of weight factors to
facilitate transitions between them. Method II and III remove one or
other of these obstacles. Specifically, method II entails the choice
of staged intermediates, but needs no weight factors, while
method III dispenses with staged intermediates, but requires
weights. The task of obtaining weights can be 
relatively time consuming (though largely intervention free if one
uses automated techniques such as the TMMC method of Appendix.~\ref{sec:tmmc}). The choice of a suitable set of stages is
somewhat less time consuming in comparison and can also be easily
automated, but is nevertheless cumbersome. This, combined with the
need in method II to perform multiple simulation at each value of $r_{\rm bb}$,
renders it slightly inferior to method III, in our view. That
said, and at the end of the day, whether one chooses to use method II
or III is probably as much a matter of personal taste than of
efficiency.

In terms of the domain of applicability of the gradual insertion approach,
methods I-III, we note that all three methods are effective in
facilitating estimates of depletion potentials at rather high volume
fractions of small particles. In this paper we have presented results
for systems having $\eta_{\rm s}^r\lesssim 0.32$ and the size ratio 
$q=0.1$. Elsewhere \cite{Ashton:2011kx} we have shown that gradual
shell insertion operates effectively up to about $\eta_{\rm
  s}^r= 0.35$. This limit arises from a rapid increase in the
relaxation time for the small particles which are tightly packed at
this volume fraction.  For size ratios smaller than $q=0.1$, the
problem of determining the depletion potential is certainly
computationally harder than for $q=0.1$ because the typical number of
overlaps $N_{\rm o}$ between the shell and the small particles is
greater. Nonetheless we still expect, in principle, to be able to reach
small particle volume fractions of $\eta_{\rm s}^r\approx 0.35$. 

Methods I-III all construct the depletion potential from measurements
of $W(r_{\rm bb})$ across a set of values of $r_{\rm
  bb}$ \footnote{Note also that methods I-III permit direct estimates
  of the contact value of the depletion potential. This contrasts with
  methods that obtain the depletion potential by integrating the
  force, which rely on extrapolation to estimate the contact
  value.}. Since each such measurement is independent, this has the
attractive feature that there are no correlations among the data
points that form the estimate of $W(r_{\rm bb})$. However, a potential
disadvantage of the approach arises from the fact that $W(r_{\rm bb})$
is obtained as the difference of two measurements, ie. $W(r_{\rm
  bb})=\ln p_i(\infty)-\ln p_i(r_{\rm bb})$. In general, both $\ln
p_i(\infty)$ and $\ln p_i(r_{\rm bb})$ are large compared to their
difference, and thus, in effect, the gradual insertion approach
calculates a small number by subtracting measurements of two large
ones. Accordingly for a given fractional uncertainty in $\ln
p_i(r_{\rm bb})$, the corresponding fractional uncertainty in
$W(r_{\rm bb})$ is larger by a factor of $\sqrt{2}\ln p_i(r_{\rm
  bb})/ W(r_{\rm bb})$, requiring a greater computational effort to
obtain a satisfactorily smooth estimate of $W(r_{\rm bb})$. To give
this issue some scale, for the case $q=0.1,\eta_{\rm s}^r=0.2$ we find
that $-\ln p_i(r_{\rm bb})\approx 200$ for shell insertion (while
incidently, $-\ln p_i(r_{\rm bb}) \approx 600$ for sphere
insertion). These are to be compared with the maximum variation in
$W(r_{\rm bb})$ of $<4$.  Increasing $\eta_{\rm s}^r$ to $0.32$ gives
for shell and sphere insertion, values of $-\ln p_i(r_{\rm bb})\approx
450$ and $-\ln p_i(r_{\rm bb})\approx 1700$ respectively, to be
compared with a maximum variation in $W(r_{\rm bb})$ of $\approx 8$.

This consideration led us to consider the utility of methods that
measure the depletion potentials by focusing on the differences in the
potential as the big particle separation is varied. Specifically we
have assessed two cluster algorithms. The GCA (method IV) is efficient
in the regime of low $\eta_{\rm s}^r\lesssim 0.2$ and represents the
method of choice in this range, delivering accurate and efficient
estimates of depletion potentials without startup costs or the need
for biased sampling.  However, to go beyond the rather limited range
of small particle volume fractions for which the GCA operates, the
constrained cluster algorithm (method V) seems a useful tool. It can
attain values of $\eta_{\rm s}^r$ as large as those accessible to the
gradual insertion methods. However, a caveat is that the apparent
smoothness of the estimates of $W(r)$ arising from method V may belie
the true absolute error in $W(r)$, which accumulates from large to
small values of $r$. Our tests in the high density regime
(cf. Fig.~\ref{fig:overall}(b)), show that for a given expenditure of
computational effort the maximum statistical error in the potential
obtained from method V is comparable, but not significantly superior
to the gradual insertions methods. However, there is scope for further
improving the efficiency of method V by using a pair of spherical caps
rather than a spherical shell for the subvolume in which preferential updating
of small particles is performed. Such subvolumes would include a
higher proportion of the small particles that are effected by the
virtual move and thus increase the rate of fluctuation in $N_o$

 \section{Summary and outlook} 

In summary, we have investigated a number of simulation techniques
that facilitate accurate measurements of depletion potentials in
highly size asymmetrical mixtures of hard spheres. Two categories of
approach were considered: (i) gradual insertion and (ii) cluster
methods.  In the first category, three flavors of methods were
described all of which obtain the depletion potential via measurements
of the insertion probability of a big sphere or shell in the
neighbourhood of another big sphere. Once prepared so that sampling
could begin, all three insertion methods showed comparable
efficiency. However, difference were found in the startup costs
associated with factors such as whether the respective methods require
precalculation of staged intermediates and/or weight factors. The
gradual insertion methods allows one to obtain depletion potentials
for small particle volume fractions of up to about $\eta_{\rm
  s}^r=0.35$. However, to reach this limit it is essential to employ
the `geometrical shortcuts' that we have described, namely shell
insertion and preferential sampling of small particles in the
neighbourhood of a big one. We remark that gradual insertion
techniques have recently been extended to systems containing many big
particles in a full grand canonical ensemble simulation scheme for
highly size asymmetrical fluid mixtures \cite{Ashton:2011fk}.

In the second category, two cluster algorithms were considered: the
Geometrical Cluster Algorithm and a bespoke constrained cluster
method. The GCA is very efficient provided $\eta_{\rm s}^r\le 0.2$.
The constrained cluster algorithm considerably extends the range of
$\eta_s^r$ for which depletion potentials can be calculated to at
least $\eta_{\rm s}^r\approx 0.32$, albeit at the price of the need to
calculate a weight function for use in biased sampling. This makes it
competitive with gradual insertion algorithms, though the hope that it
would be considerably superior in terms of overall efficiency was not
borne out due to cumulative errors.

Finally we note that with the exception of the shell trick, all the
methods considered here can be straightforwardly extended to deal with
size asymmetrical mixtures of particles interacting via more general
potentials. For the gradual insertion methods, the relevant observable
is not the number of overlaps but the energy of overlap, as 
has already described in the context of a grand canonical staged
insertion study of a highly size asymmetrical Lennard-Jones fluid
similar to method I\:\cite{Ashton:2011fk}.  For the cluster methods, a
version of the GCA suitable for arbitrary potentials is well
known~\cite{Liu2004}. The constrained cluster method could similarly be
easily extended to arbitrary interactions by considering the energy
associated with the trial move and measuring the ratio of acceptance
probabilities for the forward and reverse move.

 \acknowledgments

 This work was supported by EPSRC grants EP/I036192 and GR/F047800 and
 the Visiting Postgraduate Scholar Programme of the University of
 Bath. Some of the simulations were performed on a computer funded by
 the HEFCE infrastructure fund. VSG gratefully acknowledges the
 support of CSIC and a JAE program PhD fellowship from the Direcci\'on
 General de Investigaci\'on Cient\'{\i}fica y T\'ecnica under Grant
 No. FIS2010-15502 and from the Direcci\'on General de Universidades e
 Investigaci\'on de la Comunidad de Madrid under Grant
 No. S2009/ESP/1691 and Program MODELICO-CM. We thank Rob Jack and Bob
 Evans for useful conversations.

 \bibliographystyle{prsty}
 \bibliography{/Users/pysnbw/Dropbox/Papers}

 \appendix

 \section{Transition Matrix Monte Carlo}
 \label{sec:tmmc}

The choice of method for determining the weight function that allows
the system to sample states of low probability states is to some
extent a matter of personal taste. A number of approaches exist such
as the Wang-Landau method \cite{Wang2001a} or successive umbrella
sampling \cite{Virnau:2004bh}. In this work, we have found the
transition matrix method \cite{Smith:1995kx} to be a particularly
efficient means of determining a suitable weight function. The
transition matrix method has the attractive feature the weights can be
updated ``on the fly'' throughout the simulation, allowing the
simulation to explore an ever wider range of states as the weight
function evolves, until it eventually encompasses the state in which a
particle or shell is fully inserted.  Once this has been achieved, one
can cease updating the weight function and perform a production run
with a constant weight function.

The general idea of the transition matrix method for determining
weight functions is to record the acceptance probabilities of all
attempted transitions and extract the ratio of the states'
probabilities from it. As all attempted transitions contribute to the
weight function, including those that were rejected, the weight
function can be built up rather quickly. The details of the
implementation are summarized below, and further details have appeared
elsewhere.\cite{Errington2004,McNeil-Watson2006,Smith:1995kx}.

To implement the transition matrix method, one first defines an order
parameter $M$, for which a weight function is desired. For example in
methods III and IV, $M$ would represent the number of overlaps $N_{\rm o}$
whereas in method I it is the index $m$ of the stage. Then, for every attempted update, the
acceptance probability $p_a$ (which is calculated anyway for use in the
Metropolis criterion) is stored in a collection matrix $C$:

\begin{equation}
C(M\rightarrow M') \Rightarrow C(M\rightarrow M') + p_a\:.
\label{C_1}
\end{equation}
At the same time, the probability for rejecting the move and thereby keeping the current value of the order parameter is also stored:
\begin{equation}
C(M\rightarrow M) \Rightarrow C(M\rightarrow M) + (1-p_a)\:.
\label{C_2}
\end{equation}
It is important to note that these probabilities $p_a$ are the ``bare''
acceptance probabilities and do not include any weights. Thus for
insertions and deletions of small hard spheres, as in the present
work, they are simple zero or unity.

The transition probabilities are then calculated by normalizing the collection matrix:

\begin{equation}
T(M\rightarrow M') = \frac{C(M\rightarrow M')}{\sum_k C(M\rightarrow M_k)}\:,
\label{C_3}
\end{equation}
with the sum on the right hand side including all possible states to
which the system can jump from a given state. In the most general
case, this would create an $N \times N$ ``transition matrix'', $N$
being the number of values of the order parameter $M$ to be sampled.
The desired probability distribution $p(M)$ of the order parameter
follows as the eigenvector corresponding to the unit eigenvalue
\cite{Smith:1995kx}. However, in many case, such as the methods described in the
present paper, transition take only unit steps in the order parameter
$M$, implying that the transition matrix is tridiagonal. It follows that $p(M)$ can
be constructed simple from the ratio of the probabilities of two
adjacent values of $M$:

\begin{equation}
\frac{p(M_{i+1})}{p(M_i)} = \frac{T(M_i \rightarrow M_{i+1})}{T(M_{i+1} \rightarrow M_i)}\:,
\label{trans_simplified_1}
\end{equation}
yielding the weight difference
\bea
w(M_{i+1})-w(M_i) &=& -\ln\left(\frac{p(M_{i+1})}{p(M_i)}\right) \hfill \nonumber\\
&=&-\ln\left(\frac{T(M_i \rightarrow M_{i+1})}{T(M_{i+1} \rightarrow M_i)}\right).
\label{trans_simplified_2}
\eea

Thus, by accumulating the transition matrix in the course of a
simulation, one obtains an estimate for $P(M)$ which can be used to
update the weight function $w(M)$, thereby allowing the simulation to
explore a wider range of $M$.  Repeated updates of $w(M)$ extend
systematically the range of $M$ over which statistics
for the weight function are accumulated, until ultimately the simulation samples states in which
the big particle is fully inserted. However since updating the weight
function during a simulation violates detailed balance, we chose to do
this at rather infrequent intervals of $20000$ sweeps. 
Once the transition matrix includes value of $M$ corresponding to the
fully inserted state, the associated estimate of $p(M)$ provides a measure of the
insertion probability, as explained in Secs.~\ref{sec:methodI} and \ref{sec:methodII}.

\end{document}